\documentclass[twoside]{article}
\usepackage{multicol}
\usepackage[a4paper, margin=1.5cm]{geometry}
\usepackage{graphicx}
\usepackage{wrapfig}
\usepackage[labelformat=empty]{caption}
\usepackage{changepage}
\usepackage[table, svgnames, dvipsnames]{xcolor}
\usepackage{makecell, cellspace, caption}
\usepackage{lettrine}
\usepackage{amsmath} 
\usepackage{array}  
\usepackage[style = nature, autocite = superscript, maxbibnames = 5, url = true]{biblatex} 
\usepackage{authblk} 
\usepackage{fancyhdr}
\usepackage{hyperref}

\usepackage{mathrsfs}
\usepackage{physics}

\usepackage{ragged2e}

\AtEveryBibitem{%
  \clearfield{number}%
  \clearfield{issue}%
}
\newcolumntype{P}[1]{>{\centering\arraybackslash}p{#1}} 
\graphicspath{ {./Images/} }
\fancypagestyle{firststyle}
{
   \fancyhead[]{}
   \fancyfoot[RO]{\thepage}
   \fancyfoot[LO]{\footnotesize $^*$ Corresponding author: fluis@unizar.es}
}
\hypersetup{
    colorlinks=true,
    linkcolor=blue,
    urlcolor=blue,
    citecolor=blue
}

\begin{document}

\title{High cooperativity coupling to nuclear spins\\
on a circuit QED architecture}

\author[1,2]{V. Rollano}
\author[3]{M.C. de Ory}
\author[4]{C. D. Buch}
\author[1,2]{M. Rubín-Osanz}
\author[1,2]{D. Zueco}
\author[5]{C. Sánchez-Azqueta}
\author[6,7,8]{\\A. Chiesa}
\author[9]{D. Granados}
\author[6,7,8]{S. Carretta}
\author[3]{A. Gomez}
\author[4]{S. Piligkos}
\author[1,2,*]{F. Luis}
\affil[1]{\small Instituto de Nanociencia y Materiales de Aragón (CSIC – UNIZAR), 50009 Zaragoza, Spain}
\affil[2]{\small Departamento de F\'{\i}sica de la Materia Condensada, Universidad de Zaragoza, 50009 Zaragoza, Spain}
\affil[3]{\small Centro de Astrobiología (CSIC – INTA), Torrejón de Ardoz, 28850 Madrid, Spain}
\affil[4]{\small Department of Chemistry, University of Copenhagen, DK-2100 Copenhagen, Denmark}
\affil[5]{\small Departamento de F\'{\i}sica Aplicada, Universidad de Zaragoza, 50009 Zaragoza, Spain}
\affil[6]{\small Università di Parma, Dipartamento di Scienze Matematiche, Fisiche e Informatiche, I-43124 Parma, Italy}
\affil[7]{\small UdR Parma, INSTM, I-43124 Parma, Italy}
\affil[8]{\small INFN–Sezione di Milano-Bicocca, gruppo collegato di Parma, 43124 Parma, Italy}
\affil[9]{\small IMDEA Nanociencia, Cantoblanco, 28049 Madrid, Spain}

\date{\small \today}
\maketitle
\thispagestyle{firststyle}

\begin{center}
\parbox{14cm}{
Nuclear spins are candidates to encode qubits or qudits due to their isolation from magnetic noise and potentially long coherence times. However, their weak coupling to external stimuli makes them hard to integrate into
circuit-QED architectures (c-QED), the leading technology for solid-state quantum processors. Here, we
study the coupling of \(^{173}\)Yb(III) nuclear spin states in an [Yb(trensal)] molecule to
superconducting cavities. Experiments have been performed on magnetically diluted single crystals placed
on the inductors of lumped-element \textit{LC} superconducting resonators with characteristic frequencies
spanning the range of nuclear and electronic spin transitions. We achieve a high cooperative coupling 
to all electronic and most nuclear [\(^{173}\)Yb(trensal)] spin transitions. This result is a big leap towards the implementation of qudit protocols with molecular spins using a hybrid architecture.}

\medskip

\end{center}

\begin{multicols}{2}
\setlength{\parindent}{0em}\lettrine[lines=2,findent=0pt]{\textcolor{gray}{R}} \normalsize \hphantom ecent efforts to build the quantum computer hardware have given rise to small scale processors with more than $100$ qubits and to a staggering improvement of their operation fidelities and computing capabilities\autocite{Monz2016,Tacchino2019,Arute2019,Arute2020,Zhong2020,Wu2021,Jurcevic2021,Pogorelov2021}. Yet, reaching the conditions needed to perform large-scale computations remains a challenge for most existing platforms. The main reason is that correcting errors \autocite{Chiaverini2004,Devitt2013,Terhal2015} demands introducing a highly redundant encoding, thus significantly increasing the number of qubits needed for any  practical implementation \autocite{Fowler2012}. A promising alternative is to use $d$-dimensional qudits as the building blocks \autocite{Carretta2021}. The extra levels can help simplifying some quantum algorithms\autocite{Wang2020,Imany2019,Kiktenko2015,Lanyon2009,Kiktenko2015b,Godfrin2017} 
or quantum simulations\autocite{Tacchino2021}, ease their physical implementation and even provide a basis for embedding error correction in each single unit \autocite{Cafaro2012,Michael2016,Chiesa2020,Petiziol2021}, which fulfils both the purpose of encoding information and accounting for errors resulting from quantum fluctuations.

\setlength{\parindent}{1em} Qudits have been realized with several physical systems, including photons\autocite{Zhong2020,Arrazola2021}, trapped ions\autocite{Bruzewicz2019}
superconducting circuits\autocite{Blais2021}, and electronic and nuclear spins\autocite{Vandersypen2017,Chatterjee2021}. The latter are 
especially appealing, on account of their high degree of isolation from decoherence and the 
ability to control their spin states by means of relatively standard nuclear magnetic resonance (NMR)
techniques \autocite{Vandersypen2005}. Not surprisingly, NMR experiments performed on 
nuclear spins of organic compounds in solution provided some of the earliest implementations of quantum
protocols\autocite{Cory1998,Vandersypen2001,Ryan2009}. However, their very weak interactions with external 
stimuli also hinders 
integrating them with superconducting circuits, which form one of the most reliable platforms for solid-state 
quantum technologies\autocite{Blais2004,Xiang2013}. An interesting possibility to overcome these limitations would be to use 
nuclear spin states of magnetic ions. Indeed, the hyperfine interaction with the electronic spin mediates the coupling  of nuclear spins to electromagnetic radiation fields\autocite{Hussain2018,Gimeno2021,Chicco2021}, thereby enhancing nuclear Rabi frequencies with respect to those of isolated nuclei. 

\setlength{\parindent}{1em}Here, we apply this idea, combined with close to optimal choices of the material and the circuit design, to explore the coupling of on-chip superconducting resonators to the nuclear spins of magnetic molecules\autocite{Atzori2019,Hussain2018,GaitaArino2019,Carretta2021}. We focus on one of the simplest realizations of a molecular qudit, [Yb(trensal)]\autocite{Hussain2018, Pedersen2015,Pedersen2016}, which hosts an individual Yb(III) ion ligated by an organic molecule. This system is especially well suited to our purpose. In particular, the $^{173}$Yb isotope
exhibits an $I = 5/2$ nuclear spin coupled to an effective electronic spin $S = 1/2$. 
The strong hyperfine interaction and the presence of a sizeable quadrupolar splitting, typical of 
lanthanide ions \autocite{Bertaina2007}, give rise to the level anharmonicity necessary for
addressing the $(2S+1) \times (2I+1)=12$ spin states and, therefore, to properly encode a $d=12$ electronuclear spin qudit. In addition, the molecules form high-quality and high-symmetry single crystals in which all of them share the same axial orientation and where they can be diluted in an isostructural diamagnetic host in order to suppress decoherence. This introduces a collective enhancement of the  coupling between molecular spin excitations and microwave photons, while preserving the\unskip\parfillskip 0pt \par 


\end{multicols}

\begin{figure}[t!]
\begin{center}
\includegraphics[width = 0.935\textwidth]{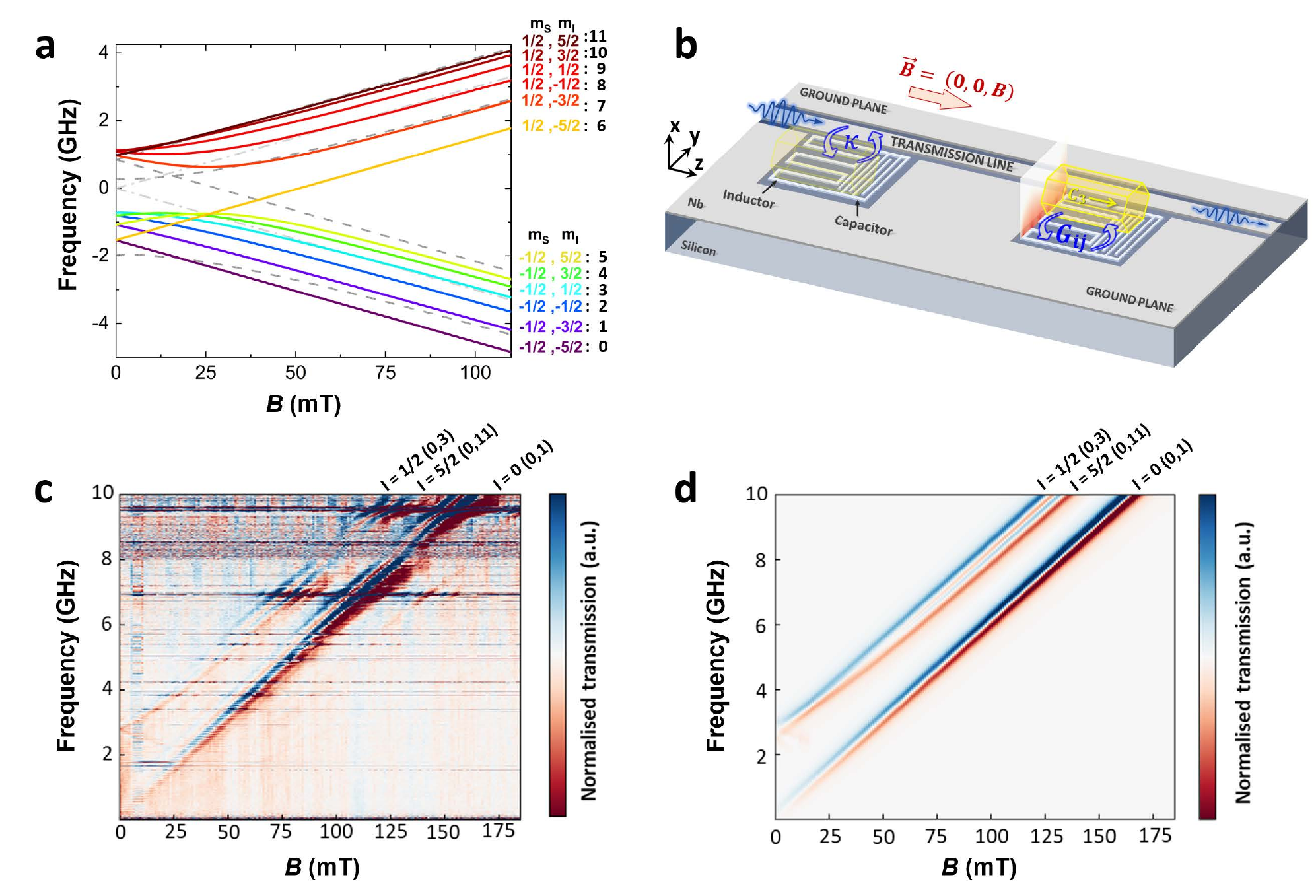}
\caption{\small {{\textbf{Fig. 1 Sample description and characterization.} \textbf{a} Spin energy levels of [\(^{{173}}\)Yb(trensal)] as a function of magnetic field. The magnetic field is parallel to the \(\textit{{C}}_{{3}}\) crystal and molecular axis, which also defines the magnetic anisotropy axis of all molecules. Each level is labelled by the effective electronic (\(\textit{{m}}_\textit{{{S}}}\)) and nuclear(\(\textit{{m}}_\textit{{{I}}}\)) spin projections that characterize the states in the high field limit. Dashed grey lines correspond to the spin energy levels of [\(^{{171}}\)Yb(trensal)], with $\textit{{I}}={1/2}$. The energy levels of all Yb isotopes with $\textit{{I}}={0}$ are shown as dashed-dotted grey lines. \textbf{b} Scheme of an [Yb(trensal)] crystal (yellow) coupled to the cavity QED system. The superconducting circuit is composed of several $\textit{{LC}}$ resonators with different characteristic frequencies, which are inductively coupled to a superconducting transmission line that reads out their respective resonances. The latter line is parallel to the external dc magnetic field and to the crystal \(\textit{{C}}_{{3}}\) axis and can also be used to perform broadband magnetic spectroscopy on a single crystal located directly onto it. \textbf{c} Experimental results of on-chip broadband spectroscopy measurements for an [Yb:Lu(trensal)] crystal with a 10\(\%\) Yb concentration. The normalised transmission \(\textit{{S}}_{{21}}\), measured at 10 mK, provides a direct map of all spin excitations as a function of magnetic field and energy (or frequency). \textbf{d} Theoretical simulation of the normalised transmission \(\textit{{S}}_{{21}}\). The conditions are the same as in (c). [Yb(trensal)] electronic spin transitions for the \(^{{171}}\)Yb, \(^{{173}}\)Yb, and other isotopes having zero nuclear spin (natural abundances 14\(\%\), 16\(\%\), and 70\(\%\), respectively) are clearly detected. Fundamental electronic spin transitions belonging to the three types of isotopes are labelled in (c) and (d).}}}
\label{fig:1}
\end{center}
\end{figure}

\begin{figure}[t!]
\begin{center}
\includegraphics[width= 0.95\textwidth]{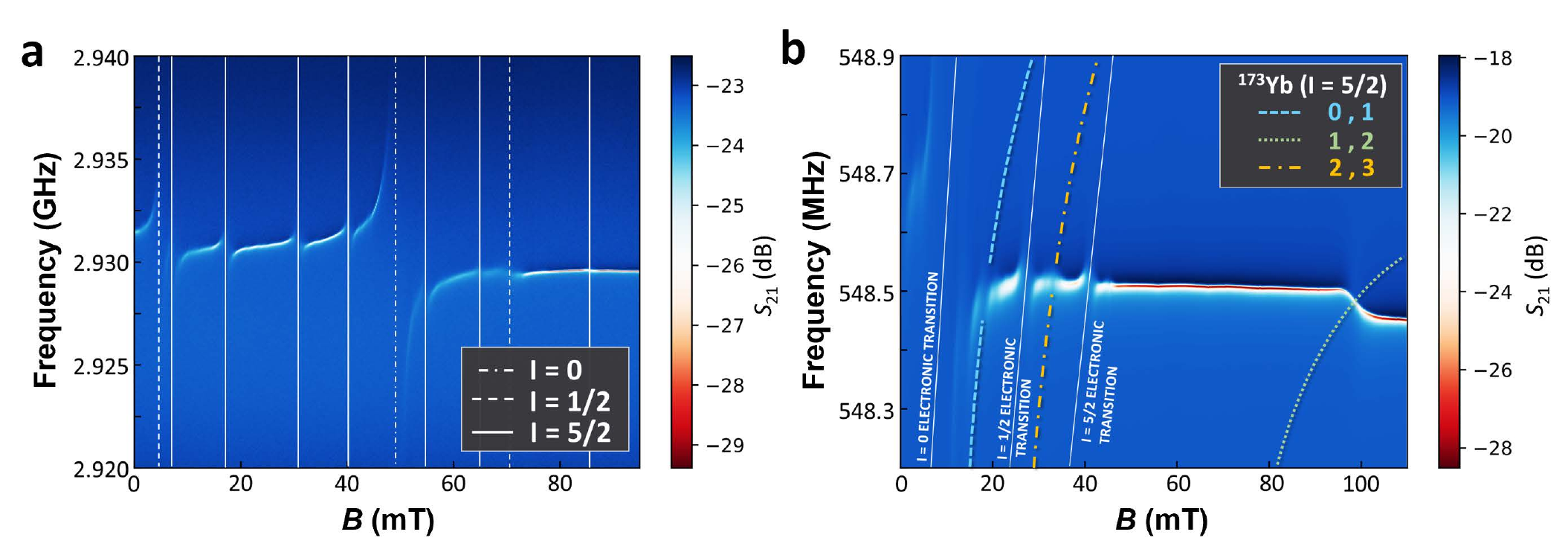}
\end{center}
\caption{{\small{\textbf{Fig. 2 Coupling cavity photons to multiple electronuclear spin transitions in [Yb(trensal)].} \textbf{a} Colour plot of the microwave transmission measured, at $\textit{T}$ = 10 mK, near the bare resonance frequency ${f}_{\rm{r}}$ = 2.93 GHz of a $\textit{{LC}}$ superconducting resonator and as a function of magnetic field. Changes in the resonator line mark the coupling of cavity photons to different electronic spin transitions. The crystal had a $\textit{{x}}$ = 5\% Yb concentration. The lines assign each of the observed features to the different Yb isotopes naturally present in this crystal \textbf{b} Same as in (a) for a ${f}_{\rm{r}}$ = 548.5 MHz resonator and a crystal with $\textit{{x}}$ = 8\%. This $\textit{{LC}}$ cavity is specifically designed to optimally couple to nuclear spin transitions in [$^{{173}}$Yb(trensal)]. Some of them are identified in the caption.}}}
\label{fig:2}
\end{figure}

\begin{multicols}{2}
\setlength{\parindent}{0em}ability to address individual spin transitions. Last, but not least, this class of molecules has been recently proposed as a suitable platform to encode error protected qubits\autocite{Hussain2018,Chiesa2020,Macaluso2020,Lockyer2021,Chicco2021,Jenkins2017}.

\setlength{\parindent}{1em}The experiments described in what follows were performed using novel, specifically designed lumped-element resonators, whose properties can be fine-tuned to match the diverse nuclear and electronic spin transitions in this molecular system \autocite{Doyle2008,Aja2021,Probst2015,Weichselbaumer2019}. For the first time, the high cooperativity regime between a nuclear spin ensemble and a superconducting resonator is achieved, with coupling strengths reaching fractions of MHz. Such coupling can be further enhanced by engineering the geometry of the resonator in order to increase the magnitude of the magnetic field in the region of the sample\autocite{Gimeno2020} and by isotopic enrichment of the crystal. Hence, these results show that the strong coupling regime for nuclear spin excitations is within reach, and represent an important step towards the realization of proof-of-concept implementations of qudit algorithms within a solid-state hybrid scheme\autocite{Jenkins2016}.

\bigskip

\setlength{\parindent}{0em}{\textbf{Results}}

\textbf{Device description.} Single crystals of [Yb(trensal)] doped into the isostructural diamagnetic host [Lu(trensal)] were obtained as described previously\autocite{Pedersen2015}. The crystals contain all stable Yb isotopes with their natural abundances and their respective nuclear spin states. The $^{173}$Yb isotope provides a natural realization of a spin qudit. The scheme of its unequally spaced, thus individually addressable, magnetic energy levels is 
shown in Fig. \hyperref[fig:1]{1a}. For not too weak magnetic fields, nuclear spin transitions (between states characterized by the same $m_S$ and $\Delta m_I = \pm 1$) have resonance frequencies below $1$ GHz, whereas electronic spin transitions (between states 
characterized by the same $m_I$ and $\Delta m_S = \pm 1$) occur at higher frequencies. Figure \hyperref[fig:1]{1a} also includes the spin levels of the $I=1/2$ and $I=0$ isotopes. 

\setlength{\parindent}{1em} The interaction of cavity photons with the molecular electronic 
and nuclear spins is realised through a superconducting circuitry
designed in the context of circuit quantum electrodynamics (c-QED) \autocite{Schuster2010, Xiang2013, Clerk2020}. 
A sketch of the device is shown in Fig. \hyperref[fig:1]{1b}. It consists of $8$ 
superconducting lumped element $LC$ resonators (LERs) coupled to 
a common readout transmission line. These LERs are designed 
with different resonance frequencies $f_{\rm{r}}$, targeting different 
electronic and nuclear spin transitions. Since the LERs are 
parallel coupled to the transmission line, parameters such as resonance frequency and impedance 
can be freely designed without affecting the 
transmission through the device. Thus, it is possible to tailor the microwave field 
$\vec{b}(\vec{r})$ generated by the inductor line in order to optimize 
the coupling rate per spin within the crystal volume. A contour plot of the field created by one 
of these resonators is included in Fig. \hyperref[fig:1]{1b}, next to the crystal and LER 
located on the right hand side (see also Fig. \hyperref[fig:S1]{S1} of the Supplemental Information). 

\setlength{\parindent}{1em}Different [Yb:Lu(trensal)] crystals, with concentrations $x$ ranging 
from $2$\% to $8$ \%, were placed on top of each of the resonators. The   
magnetic anisotropy axes lie parallel to the transmission line and to the external dc magnetic field $\vec{B}$. In this geometry, $\vec{b}(\vec{r})$ gives rise to finite transition rates between multiple 
pairs of spin states linked by nonzero matrix elements, as far as the spin energy levels are brought to 
resonance with the LER by the action of $\vec{B}$. Experiments have been performed with an input driving power of $-95$ dBm. This corresponds to a microwave photon number $\sim 10^{8}$ much lower than the number of spins involved in the coupling ($\sim 10^{16}$), thus avoiding power saturation effects (experiments aiming the optimization of the driving power are shown in Fig. \hyperref[fig:S2]{S2}).

\setlength{\parindent}{0em}\textbf{Broadband spectroscopy.} In order to identify the origin of the 
different resonances and to guide the design of the resonators, it is important to characterize the energy spectrum of 
the different Yb isotopes under the same conditions. To this aim, we performed experiments on a $x=10$\% single crystal 
directly coupled to the superconducting readout transmission line (see Fig. \hyperref[fig:S3]{S3} for a sketch of this experimental configuration 
and additional results). These experiments allow us to address spin transitions at any possible driving frequency $< 14$ GHz. 
The normalized transmission amplitude $|S_{21}(B,f)|$ as a function of external magnetic field and driving frequency is shown in Fig. \hyperref[fig:1]{1c}. Details on the normalization of the experimental data are described in the Supplemental Information. The results 
provide a full picture of the Zeeman energy diagram. Resonances arising from the coupling to electronic spin transitions in $^{173}$Yb ($I = 5/2$) and $^{171}$Yb ($I = 1/2$), and to isotopes with zero nuclear spin are clearly visible. The comparison with 
theoretical simulations (see Fig. \hyperref[fig:1]{1d} and methods) reveals that the ratios between the intensities of these signals are $14\%$ $(^{171}$Yb), $16\%$ $(^{173}$Yb), and $70 \%$ (all $I=0$ isotopes), in good agreement with their natural abundances. The half-width $\gamma$ of the electronic transitions is $\sim 23$ MHz. This parameter characterizes the inhomogeneous broadening and it is the main limiting factor to reach a coherent coupling of the spins to the superconducting cavities \autocite{Nellutla2007,Schuster2010}.

\setlength{\parindent}{0em}\textbf{Coupling cavity photons to multiple electronuclear spin transitions.} The complex spin level spectrum of [$^{173}$Yb(trensal)] and the presence of different
Yb isotopes lead to multiple resonances of spin transitions ${i,j}$ with each of the resonators 
(here the $i$ and $j$ indexes label the spin levels in increasing energy, from $0$ up to the highest excited level of each isotope). 
Illustrative examples are shown in Fig. \hyperref[fig:2]{2}. The microwave transmission $|S_{21}|$ measured as a function of
magnetic field and driving frequency at $10$ mK shows clear signatures of the coupling to the 
spins, namely a shift of the cavity frequency and a drop in its visibility. The high quality factor of these resonators leads to bare line widths $\kappa$ of the order of a few kHz, which 
contribute to a very high sensitivity and allow resolving each of the individual transitions at their respective resonant fields. For frequencies above $1$ GHz ($f_{\rm r} = 2.93$ GHz in Fig. \hyperref[fig:2]{2a}), only electronic spin transitions contribute. Using the information about the field dependence of these spin levels obtained previously (Fig. \hyperref[fig:1]{1}), it is possible to assign all these features to specific electronic spin transitions of the different isotopes. As with the broadband spectroscopy results discussed above, the coupling to the more abundant $I=0$ isotopes shows the highest visibility. This agrees with the fact that the visibility of any spin resonance ${i,j}$ increases with the collective spin-photon coupling $G_{{i,j}}$, which is proportional to $\sqrt{N_{i,j}}$, where $N_{i,j}$ is the number of spins involved in each particular transition. 

\setlength{\parindent}{1em}For frequencies below $1$ GHz, nuclear spin transitions from the isotopes having nonzero $I$ are also observed. In particular, Fig. \hyperref[fig:2]{2b} shows the coupling of several $^{173}$Yb nuclear spin transitions to a $f_{\rm r} = 548.5$ MHz superconducting cavity. These transitions are identified by their resonant magnetic fields and weaker visibility as compared to the electronic ones. Combining experiments performed on multiple resonators having different frequencies (in our case, $f_{\rm r} = 403.6$ MHz, $414.3$ MHz, $548.5$ MHz and $549.5$ MHz) allows tracking how several of these transitions evolve with magnetic field. These data are shown in Fig. \hyperref[fig:S4]{S4}. Nuclear spin transitions are characterized by a much weaker magnetic field dependence, as expected (see Fig. \hyperref[fig:1]{1a}). 

\begin{figure}[b!]
\centering
\includegraphics[width=0.95\linewidth]{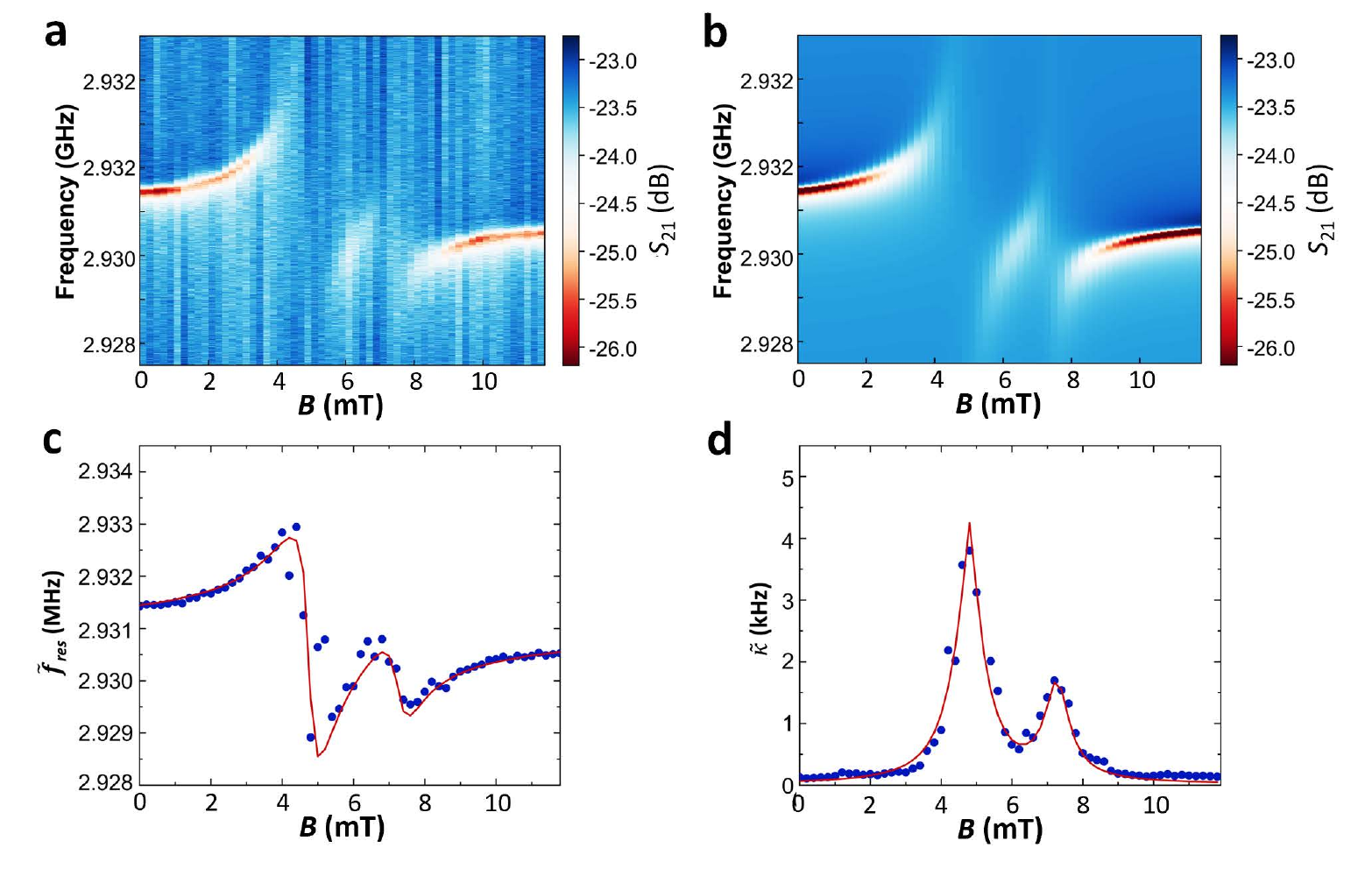}
\captionof{figure}{{\small{\textbf{Fig. 3 Coupling cavity photons to the electronic spin.} \textbf{a} Colour plot of the microwave transmission measured, at $\textit{{T}}$ = 10 mK, near the bare resonance frequency ${f}_{\rm{r}}$ = 2.93 GHz of a superconducting LER. The magnetic field brings the splittings between levels 0 and 3 of $^{{171}}$Yb and $0$ and $11$ of $^{{173}}$Yb into resonance with the cavity photons, leading to the observed avoided crossings. \textbf{b} Optimized  simulation of the experimental data shown in (a), from which the collective spin-photon coupling and the spin line width are determined. \textbf{c} Resonance frequency (\(\tilde{f}_{\textsf{r}}\)) and \textbf{d} line width (\(\tilde{\kappa}\)) of the coupled spin-LER system  as a function of magnetic field, extracted from the experimental data (blue dots) and from the fit (solid red line).}}}
\label{fig:3}
\end{figure}

\setlength{\parindent}{0em}\textbf{High cooperative coupling to electronic spins.} Each of the observed transitions ${i,j}$ can then be analyzed in detail, in order to determine the strength of 
the spin-photon coupling $G_{{ij}}$ and the spin resonance line widths $\gamma_{i,j}$. We first 
consider electronic spins. Figure \hyperref[fig:3]{3a} shows an illustrative example, taken from the low field region of Fig. \hyperref[fig:2]{2a}. This LER couples to the ${0,3}$ and ${0,11}$ electronic spin transitions of, respectively, $^{171}$Yb ($I=1/2$) and $^{173}$Yb (our $I = 5/2$ qudit). A $2D$ least-squares fit of these experimental data, based on a generalized input-output model described in the methods section below, is shown in Fig. \hyperref[fig:3]{3b}. As  can be seen in Figs. \hyperref[fig:3]{3c} and \hyperref[fig:3]{3d}, the fit reproduces very well the effect that the coupling to the spins has on the effective LER frequency $\tilde{f_{\rm r}}$ and resonance width $\tilde{\kappa}$, thus showing that this model provides a reliable description of the underlying physics. 

\setlength{\parindent}{1em}The fit gives $G_{0,3} = 7.9$ MHz for $^{171}$Yb and $G_{0,11} = 5$ MHz for $^{173}$Yb. Values for other electronic transitions are shown in Tables \hyperref[tab:S1]{S1} and \hyperref[tab:S2]{S2}. The line widths are in good agreement with those derived from broadband spectroscopy experiments. They decrease with concentration, from about $22$ MHz for $x=8$ \% to $13$ MHz for $x=2$\%. The limit of strong coupling, defined by the condition $G_{i,j} > \kappa$ and $\gamma_{i,j}$ is only achieved for the $I=0$ isotopes (see Fig. \hyperref[fig:2]{2a}, Fig. \hyperref[fig:S5]{S5} and Table \hyperref[tab:S1]{S1}) for which $G_{0,1} = 22$ MHz, again as a consequence of their larger natural abundance. Still, all transitions measured reach the high cooperative coupling, characterized by a larger than unity cooperativity parameter $C_{i,j} \equiv {(G_{i,j})^2}/{(\gamma_{i,j} \cdot \kappa)}$. For the transitions shown in Fig. \hyperref[fig:3]{3}, $C_{0,3} = 129$ and $C_{0,11} = 46$, while for the $I=0$ isotopes $C_{0,1}$ reaches $946$. Reaching this limit implies that, at resonance, nearly every photon in the cavity is coherently transferred to the spin ensemble \autocite{Schuster2010}.

\end{multicols}

\newpage

\begin{figure}[b!]
\begin{center}
\includegraphics[width=0.94\linewidth]{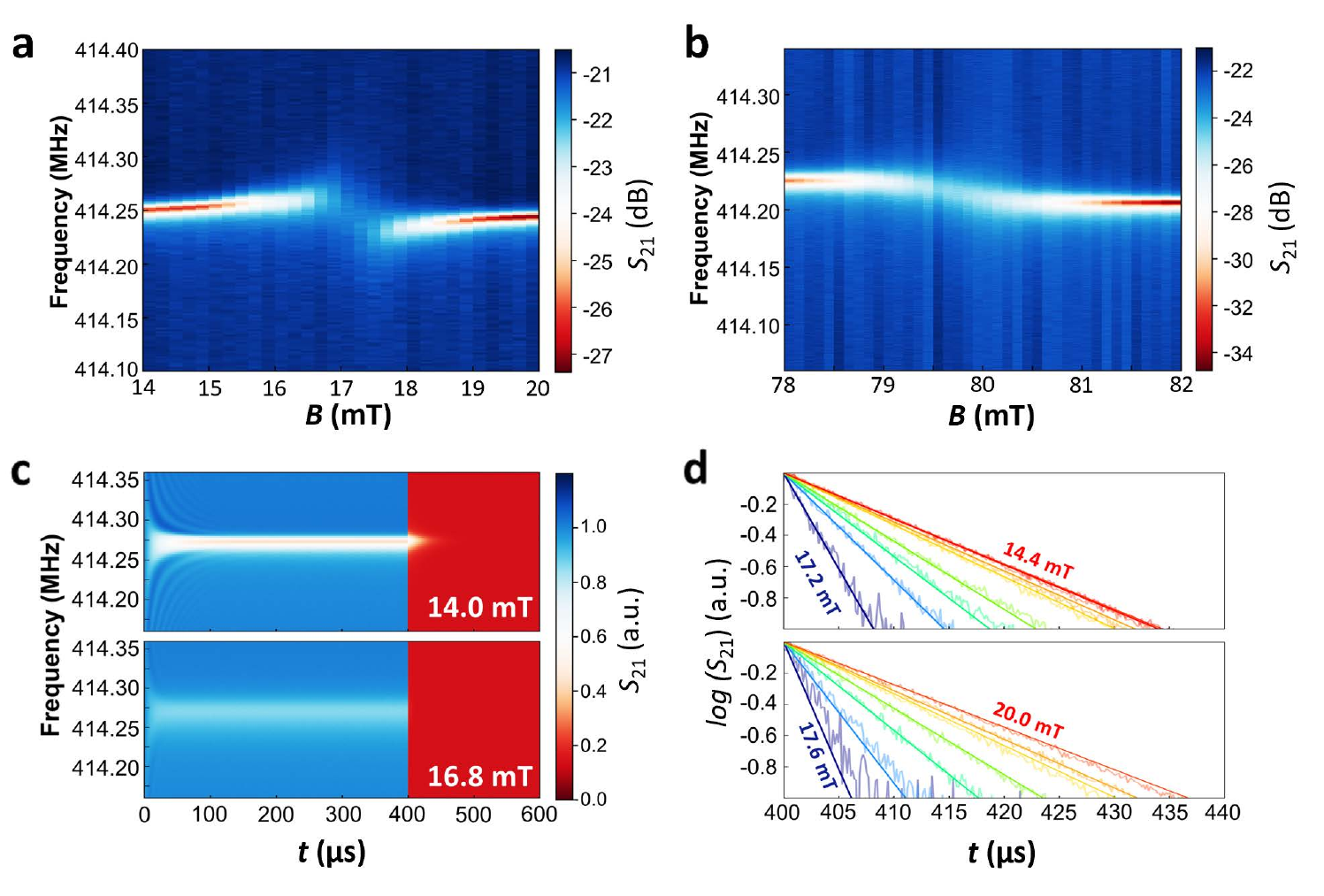}
\captionof{figure}{{\small{\textbf{Fig. 4 Coupling cavity photons to nuclear spins.} \textbf{a} Colour plot of the microwave transmission showing the resonance of a \(f_r\) = 414 MHz superconducting LER to the ${1,2}$ nuclear spin transition in [\(^{{173}}\)Yb(trensal)]. \textbf{b} Same as in (a) for the ${2,3}$ nuclear spin transition. \textbf{c} Time and frequency dependence of the transmission measured near the resonance of a \(f_r\) = 414 MHz LER following the application and removal of a $400$ $\mu$s long pulse through the readout transmission line. The two plots show the behaviour observed in and out of resonance with the ${1,2}$ nuclear spin transition in [\(^{{173}}\)Yb(trensal)]. Time dependent curves measured at constant driving frequencies are shown in Fig. \hyperref[fig:S11]{S11}. \textbf{d} Time dependence of the resonator discharge following the end of the applied microwave pulse. The top and bottom panel show data measured for dc magnetic fields increasing towards and beyond the resonance between spins and photons, respectively. The dc magnetic field step is 0.4 mT for both.}}}
\label{fig:4}
\end{center}
\end{figure}

\begin{multicols}{2}
\setlength{\parindent}{0em}\textbf{High cooperative coupling to nuclear spins.} Reaching a similar condition with the nuclear spin states is, a priory, much more demanding on account of the very small nuclear magnetic moments. However, as we have mentioned above, the electronic spin introduces a quite efficient path to couple the nuclear spins to external magnetic fields. Therefore, it is to be expected that the same effect also enhances the coupling to cavity photons. 

\setlength{\parindent}{1em}Figures \hyperref[fig:4]{4a} and \hyperref[fig:4]{4b} show experimental data for the ${1,2}$ and ${2,3}$ nuclear spin transitions in $^{173}$Yb(trensal), which are brought into resonance with a $f_{\rm{r}} = 414.3$ MHz LER near $17$ mT and $80$ mT, respectively. The highest coupling rates for these transitions, $G_{\rm{1,2}} = 0.24$ MHz and $G_{\rm{2,3}} = 0.11$ MHz, are achieved at $10$ mK and $50$ mK, respectively. Other nuclear spin transitions give comparable results (see Table \hyperref[tab:S3]{S3} and Figs. \hyperref[fig:S6]{S6} and \hyperref[fig:S7]{S7} for the complete set). For instance, the ${0,1}$ transition resonates with a $548.5$ MHz LER at about $20$ mT, as shown by Fig. \hyperref[fig:2]{2b}, with $G_{0,1} = 0.13$ MHz. The corresponding line widths are $\gamma_{\rm{0,1}} = 2.1$ MHz, $\gamma_{\rm{1,2}} = 1.6$ MHz and $\gamma_{2,3} = 0.6$ MHz, consistently with the effect of strain on quadrupole and hyperfine parameters. Indeed, the hierarchy between the $\gamma_{i,j}$ reflects the one between the derivatives of the corresponding gaps with respect to the Hamiltonian parameters $p$, $A_{x,z}$ and $H$ (see Figs. \hyperref[fig:S8]{S8} and \hyperref[fig:S9]{S9}), pointing to inhomogeneous broadening as the origin of the observed line width. 
With a resonator half-width $\kappa \simeq 3.78$ kHz, the high cooperativity limit is achieved for all of them: $C_{0,1} = 2.0$, $C_{1,2} = 9.1$ and $C_{2,3} = 4.9$. The highest cooperativity has been achieved for the ${1,2}$ resonance at $f_{\rm{r}} = 403.3$ MHz and a relatively high Yb concentration $x = 8 \%$, for which $G_{1,2} = 0.43$ MHz, $\gamma_{1,2} = 2.3$ MHz, and $C_{1,2} = 24.0$, thus close to values obtained for the electronic spin transitions. 

\setlength{\parindent}{1em}Although the couplings and the cooperativity are about ten times smaller than what is achieved for the electronic transitions in the same molecular system, they are nevertheless remarkably high. If the nuclear and electronic spins were uncoupled, i.e. if the hyperfine interaction vanished, the nuclear spin photon coupling would be mediated solely by the nuclear Zeeman interaction. As it shown in Fig. \hyperref[fig:S10]{S10}, this would lead to a ratio of about $10^{-4}$ between the coupling rates of nuclear and electronic transitions (say ${1,2}$ vs ${0,11}$). Therefore, the electronic spins help to mediate a much stronger interaction with the circuit photons.

\setlength{\parindent}{0em}\textbf{Coupling rates temperature dependence.} Following the same procedure, we have determined the coupling constants of most detectable transitions as a function of temperature, which controls the relative populations\unskip\parfillskip 0pt \par 

\begin{center}
\includegraphics[width=0.9\linewidth]{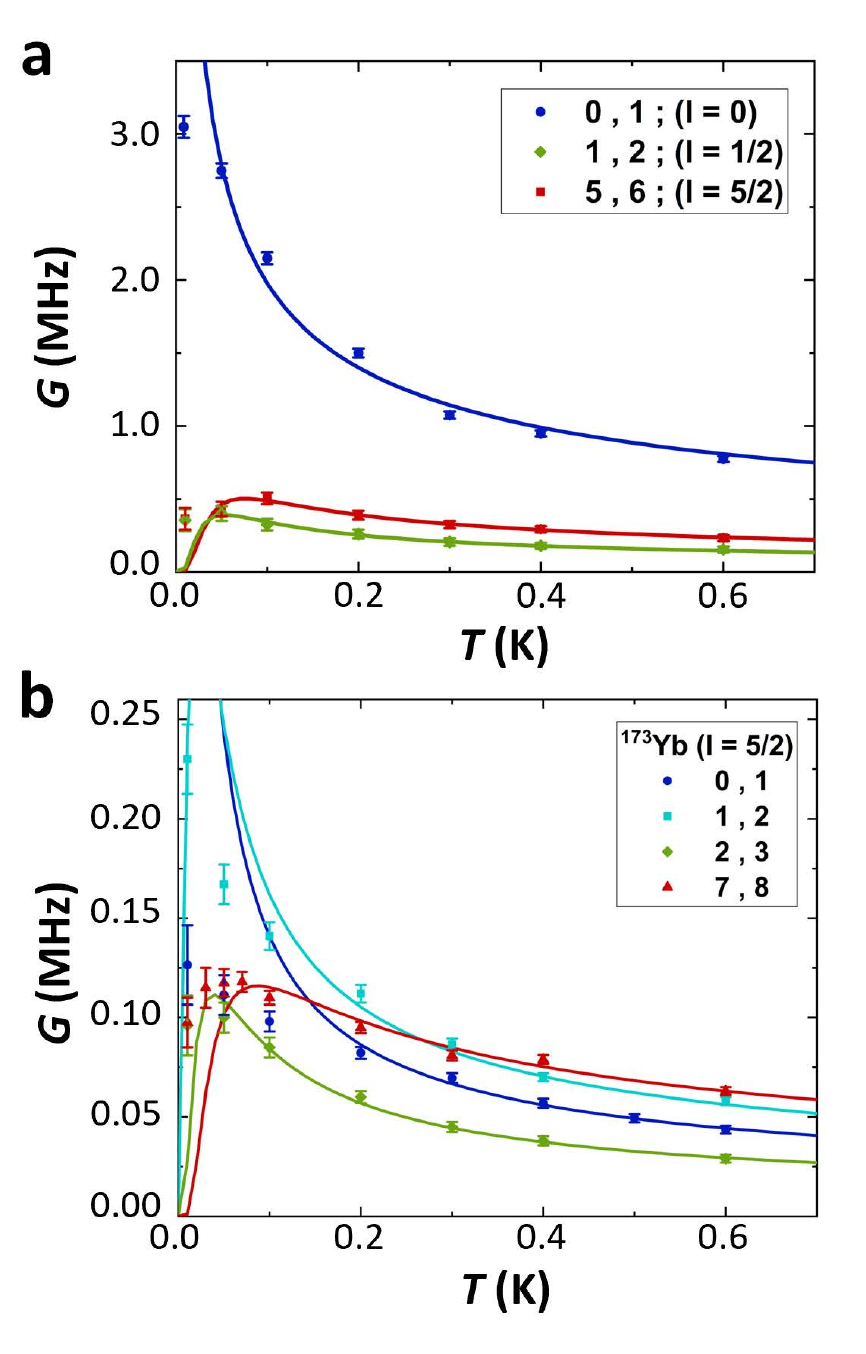}
\captionof{figure}{{\small{\textbf{Fig. 5 Evolution of the spin-photon rates with temperature.} \textbf{a} Experimental (dots) and theoretical (lines) coupling rates as a function of temperature for the three
electronic spin transitions coupled to a $f_{\rm{r}} =$ 414.3 MHz cavity. \textbf{b} Experimental (dots) 
and theoretical (lines) coupling rates as a function of temperature for different nuclear
spin transitions in [$^{{173}}$Yb(trensal)]. Here, the ${0,1}$
transition resonates at $f_{\rm r} =$ 548.5 MHz, while the others are measured with the same 414.3 MHz cavity as the electronic spin transitions shown in (a).}}}
\label{fig:5}
\end{center}

\medskip

\setlength{\parindent}{0em}of the electronuclear spin levels and, therefore, also the number of spins involved in the coupling to photons. Results obtained for electronic and nuclear spin transitions coupled to $\sim 0.5$ GHz LERs are shown in Fig. \hyperref[fig:5]{5} (see also Figs. \hyperref[fig:S6]{S6} and \hyperref[fig:S7]{S7}). They are compared to numerical simulations based on the spin Hamiltonian of each species and on high resolution $3D$ maps of the magnetic field generated by the LER (Figs. \hyperref[fig:1]{1b} and \hyperref[fig:S1]{S1}). The only fitting parameter is an overall scaling factor that parameterizes the effective filling of the LER mode by the molecular crystal. The theory (see methods) reproduces fairly well the dependence of $G_{i,j}$ with $T$, although deviations are seen at the lowest temperatures. These mark the point where the relative spin level populations involved in each transition deviate from equilibrium. The fact that this happens at higher temperatures (of order $200$ mK) for the nuclear spins than for the electronic spins ($\leq 20$ mK), suggests that the former have longer spin-lattice relaxation times $T_{1}$. Together with the smaller line widths observed for them ($\sim 1-2$ MHz vs $\sim 14-16$ MHz for the electronic ones), this confirms that nuclear spins in [$^{173}$Yb(trensal)] remain less sensitive than the electronic ones to external perturbations and decoherence. 

\setlength{\parindent}{0em}\textbf{Time-resolved experiments.} We have further investigated this point by means of time-resolved experiments, recording the charging and discharging of the resonator following the application of square microwave pulses through the readout line. Results of transmission as a function of driving frequency and time are presented in Fig. \hyperref[fig:4]{4c} at two different magnetic fields corresponding to in and off resonant conditions with the ${1,2}$ nuclear spin transition of [$^{173}$Yb(trensal)]. Additional data are shown in Fig. \hyperref[fig:S11]{S11}. In the stationary state, attained near the end of the pulse duration, the response agrees perfectly well with results measured with the continuous wave method. Monitoring the response at short times after switching on and off the pulse allows characterizing the lifetime of the cavity photons and how it is affected  by their coupling to either electronic or nuclear spins. This decay is close to exponential, $|S_{21}| \simeq e^{-\tilde{\kappa} \cdot t}$, as shown in Fig. \hyperref[fig:4]{4d}, and becomes faster whenever the photons become hybridized with the spins. Besides, the characteristic time constant $T_{2}^{*}$ obtained near resonance is about ten times longer for nuclear spin transitions than for electronic ones, at the same frequency, and it agrees well with the $1/\gamma$ value derived from continuous wave measurements.

\bigskip

\setlength{\parindent}{0em}{\textbf{\normalsize Discussion}}

\setlength{\parindent}{0em} We have shown for the first time a high-cooperativity regime between nuclear spins and photons in an on-chip superconducting resonator setup. In particular, we have achieved this regime for both electronic and nuclear spin excitations in crystals of [Yb(trensal)] molecules. The hyperfine interaction with the electronic spin of the same ion plays here a key role as it enhances the interaction of nuclear spins to microwave magnetic fields, thus allowing them to reach the high cooperativity regime without compromising much their good isolation from noise\autocite{Hussain2018,Chicco2021}. The balance between spin operation rates and coherence can furthermore be tuned by the magnetic field, which controls the degree of electronuclear spin entanglement \autocite{Gimeno2021}. The results also exemplify the application of these devices to perform magnetic resonance at diverse frequencies and in a single experimental run, from which a complete characterization of multiple spin levels of different species (in our case, isotopical analogues of the same molecule) can be obtained. 

\setlength{\parindent}{1em} Attaining a sufficiently high coupling of cavity photons to spin qubits and qudits is also a crucial ingredient to use them in the context of quantum technologies, either as quantum memories\autocite{Wesenberg2009,Julsgaard2013} or as operational units of a hybrid processor\autocite{Carretta2013}. Even without reaching strong coupling, the circuit used in this work can serve as a basis for proof-of-concept realizations (on ensembles) of algorithms based on a qudit encoding. As proposed in \cite{Hussain2018,Chiesa2020,Petiziol2021}, a nuclear spin $I>1$ coupled to an electronic effective spin $1/2$ (of which [Yb(trensal)] is a prototypical realization) can be exploited to embed an error-protected logical unit within a single molecule. In the proposed setup, the control of nuclear spins, which encode the error protected unit, can be done by introducing an extra excitation line and using conventional NMR techniques \autocite{Vandersypen2005}. Then, a sufficiently high cooperativity would provide a way to dispersively read out the results \autocite{Gomez2022}. 

Our results show that we are not far from reaching this limit. Besides, they provide ample room for improving the spin photon couplings. First, one can work with isotopically pure crystals containing only the most interesting [$^{173}$Yb(trensal)] molecules. For any concentration $x$, this introduces a collective enhancement of the coupling by a factor $\sqrt{1/0.16} \simeq 2.5$ with respect to samples with natural isotopical abundances. Next, the interface between the sample and the crystal also plays a very important role \autocite{Urtizberea2020}. The comparison of the measured coupling rates to theoretical simulations suggest that the filling factor of the cavity modes is as low as $4 \%$, which is compatible with the existence of a gap of about $50$  $\mu m$ between the chip surface and the crystal (Fig. \hyperref[fig:S12]{S12}). This leaves margin for a further factor $5$ enhancement in $G_{i,j}$, perhaps even more, by exploiting the possibilities to design LERs adapted to each situation. For instance, one could work with smaller crystals coupled to resonators having correspondingly smaller inductor lines or, the opposite, using resonator designs leading to larger cavity volumes. Then, even achieving the strong coupling regime seems feasible. In this regime, we expect $\gamma$ to be further reduced, thanks to the cavity-protection mechanism\autocite{Diniz2011,Putz2014,Chiesa2016}. The strong coupling of an inhomogeneously broadened spin ensemble with the resonator introduces a gap between the bright and dark (subradiant) states, which protects the former from decay. This condition is met if the distribution of the spin ensemble frequencies decays faster than a lorentzian, as we expect here.

Combined with suitable circuit designs \autocite{Bienfait2016,Probst2017} and including nanometer wide inductor lines in order to confine and locally enhance the microwave magnetic field \autocite{Gimeno2020}, these ideas might even allow reaching sizeable couplings to individual nuclear spins, thus opening the perspective of wiring up different nuclear spin qudits into a scalable architecture \autocite{Carretta2021}.  In summary, the results offer many promising prospects for the integration and exploitation of nuclear spins in circuit-QED quantum architectures.

\bigskip

\setlength{\parindent}{0em} {\textbf{\normalsize Methods}}
\smallskip

\setlength{\parindent}{0em}\textbf{Experimental details.} The superconducting cavities are based on a lumped-element resonators (LERs) design \autocite{Aja2021}. They are fabricated by maskless lithography and reactive ion etching techniques on a $100$ nm thick Nb film deposited 
by means of DC magnetron sputtering on a $350$ \(\mu\)m thick silicon substrate. The base pressure prior to the Nb deposition is better than \(2.0 \times 10^{-8}\) Torr. The cavities were designed using the commercial software Sonnet for the RF simulations and our CUDA code to calculate \(\vec{b}(\vec{r})\) (see Fig. \hyperref[fig:S1]{S1}). 

\setlength{\parindent}{1em}Single-crystals of [Lu:Yb(trensal)] (where H\(_{3}\)trensal = 2,2,2-tris(salicylideneimino)-triethylamin) employed in these experiments consist of [Yb(trensal)] doped into the isostructural [Lu(trensal)] at different concentrations $x$. They were synthesized and grown over several weeks following a method described previously for Er(trensal)\autocite{Pedersen2014}, which employed a mixture of Yb(CF$_3$SO$_3$)$_3$·9H$_2$O and Lu(CF$_3$SO$_3$)$_3$·9H$_2$O. The crystal and molecular structures were characterized by X-ray diffraction of single-crystals and powder samples (crashed single-crystals), verifying their nature and purity. The complexes crystallize in the trigonal P\(\bar{3}\)C1 space group. The actual Yb concentration was determined by Inductively Coupled Plasma Mass Spectrometry (ICP-MS, Bruker Aurora Elite). The instrument was tuned prior to measurement and external calibration was performed using calibration points spanning the range of concentrations encountered in the samples. The nitric acid used was TraceSelect grade, and the reference material was provided by Inorganic Ventures. The sample was prepared by dissolving a single crystal from the same crystallization batch as the one providing the crystal for the experiments presented in 2\(\%\) nitric acid (dissolved in 50 mL from which 0.125 mL was diluted to 50 mL). The sample preparation for ICP-MS and its analysis were performed at the Department of Chemistry, University of Copenhagen. Finally, the crystals were located on top of their respective superconducting LER. To improve the crystal-chip interface, we covered and gently pressed them with a Teflon tape, fixed to the ground planes with low temperature grease (Apiezon N). The chip was mounted in a BlueFors LD450 dilution refrigerator equipped with a 1 T superconducting magnet. A cold finger is used to place the chip in the centre of the magnet. As shown in Fig. \hyperref[fig:S1]{S1}, the microwave magnetic field created by the cavities is mostly perpendicular to the crystal anisotropy axis. The geometry of the setup is shown in Figs. \hyperref[fig:1]{1b} and \hyperref[fig:S3]{S3}. The chip is connected through coaxial lines to a Vector Network Analyzer (VNA) with a measurement frequency bandwidth ranging from 10 MHz to 14 GHz. The device is then characterized by measuring the transmitted signal \(S_{21}\). 

\medskip

\small \setlength{\parindent}{0em}\textbf{Spin Hamiltonian.} In our calculations, [Yb(trensal)] is described by the effective spin Hamiltonian: \\
\begin{equation} \label{eq:1} \begin{split} \mathcal{H} = \mu_{\rm{B}} \vec{S} \cdot {g}_{\rm{e}} \cdot \vec{B} - \mu_{\rm{N}} g_{I} \vec{I} \cdot \vec{B} + \phantom{{}={}} \\
\phantom{{}={}} + A_{\parallel} S_{z}I_{z} + A_{\perp}(S_{x}I_{y} + S_{y}I_{y}) + p I_{z}^2
\end{split}
\end{equation}

\smallskip

\normalsize The first two terms correspond to the Zeeman coupling of 
electronic and nuclear spins, respectively, to the magnetic field. The \(A_\parallel\) and \(A_\perp\) terms describe the hyperfine coupling while the last term accounts for the quadrupolar interaction of the nuclear spin. Parameters for the three different Yb isotopes are summarized in Table \hyperref[tab:1]{1}.

\begin{center}
\renewcommand{\arraystretch}{1.40}

\begin{tabular}{|P{0.85cm}|P{0.65cm}|P{0.65cm}|P{1.05cm}|P{0.8cm}|P{0.8cm}|P{0.85cm}|}
    \hline
    \rowcolor{black}
    
    \textcolor{white}{} &  \footnotesize\centering \textcolor{white}{\(g_{e\perp}\)} & \footnotesize\centering \textcolor{white}{\(g_{e\parallel}\)} & \footnotesize\centering \textcolor{white}{\(g_{I}\)} & \scriptsize\centering\textcolor{white}{\(A_{\perp}\)\newline(GHz)} & \scriptsize\centering\textcolor{white}{\(A_{\parallel}\)\newline(GHz)} & \scriptsize\textcolor{white}{\(p\)\newline(GHz)} \\
    \footnotesize \textbf{\(^{171}\)Yb} & \scriptsize 2.935 & \scriptsize 4.225 & \scriptsize -0.02592 & \scriptsize 3.3729 & \scriptsize 2.2221 & \scriptsize 0\\ 
    \footnotesize \textbf{\(^{I=0}\)Yb} & \scriptsize 2.935 & \scriptsize 4.225 & \scriptsize 0 & \scriptsize 0 & \scriptsize 0 & \scriptsize 0\\
    \footnotesize \textbf{\(^{173}\)Yb} & \scriptsize 2.935 & \scriptsize 4.225 & \scriptsize -0.02592 & \scriptsize -0.897 & \scriptsize -0.615 & \scriptsize -0.066\\
    \hline
\end{tabular}
\captionof{table}{{\small {\textbf{Table 1. Parameters of the effective spin Hamiltonian for different [Yb(trensal)] isotopical derivatives.}}}}
\label{tab:1}
\end{center}

\textbf{Coupling of spins to a transmission line.} The broadband spectroscopy experiments were analysed using an input-output theory for a $1D$ coplanar transmission line adapted to the conditions of our experiments. In this model, the coupling \(\Gamma_{i,j}\) of a particular spin transition with resonance frequency \((\Omega_{i,j})\) to the photons with frequency \(f\) travelling through the transmission line is given by:

\begin{equation} \label{eq:2}
\Gamma_{i,j} = 2 \pi g \cdot \big| \langle \psi_j | V | \psi_i \rangle \big|^2 \cdot \Delta P_{i,j} \cdot \Omega_{i,j} \cdot (n_{i,j} + 1)
\end{equation}

\smallskip

Where $g(f)$ is a spin-photon coupling density, which depends on the mode density in
the transmission line and on a geometrical factor. The operator \(V(\vec{r}) = g_{\rm{e}} \mu_{\rm{B}} \vec{b}(\vec{r}) \cdot \vec{S} - g_{\rm{I}} \mu_{\rm{N}} \vec{b}(\vec{r}) \cdot \vec{I}\) mediates the interaction with the photons in the line at the position $\vec{r}$, \(\Delta P_{i,j}\) is the population difference between spin states \(i,j\) involved in the transition, and \(n_{i,j}\) is the bosonic occupation number. Therefore, the transmission through the superconducting line is modelled as: 

\begin{equation} \label{eq:3} 
S_{21} = \frac{\alpha}{1 + \sum_{i,j}\frac{\Gamma_{i,j}}{\gamma_{i,j} + i(\Omega_{i,j} - f)}}
\end{equation}

\smallskip

Where \(\alpha\) is a complex attenuation factor accounting for the dependence of the transmission line losses as a function of the driving frequency.\\

\textbf{Spin-resonator coupling.} 
For the resonator-spin case the transmission probes the coupled system via the $LC$-resonator. The simplest model considers the two resonance frequencies, one for the spins $\Omega_{i,j}$ and another one for the $LC$-resonator, $f_{\rm r}$.  The coupling between them is $G_{i,j}$. In this case $S_{21}$ is computed by solving the set of coupled equations 
\smallskip
\begin{equation} \label{eq:4}
\footnotesize \begin{pmatrix} 
i(f_{\rm{r}} - f) + \kappa & iG_{i,j} \\
iG_{i,j} & i(\Omega_{i,j} - f) + \gamma_{i,j}
\end{pmatrix} \cdot \begin{pmatrix} b_1 \\ b_2 \end{pmatrix} = - i\sqrt{\kappa_{e}e^{i\varphi}} \begin{pmatrix}  a_{1}\\ a_{2} \end{pmatrix}
\end{equation}
\smallskip

and making \(S_{21} = \alpha \cdot |1 -  i\sqrt{\kappa_{e}e^{i\varphi}} \cdot b_{1}|\). Notice that since the system is driven through the superconducting cavity, \(a_1\) = 1 and \(a_2\) = 0. Actually, it is possible to obtain Eq. (\hyperref[eq:3]{3}) from Eq. (\hyperref[eq:4]{4}) with $G_{i,j}$ = 0 making $\kappa_{e} = \Gamma_{i,j}$ and $\kappa - \kappa_{e}e^{i\varphi} = \gamma_{i,j}$ (thus replacing the resonator with the spin ensemble). 

\setlength{\parindent}{1em}The diagonal elements of the interaction matrix include the properties (resonance frequency and losses) of two oscillators: the superconducting LER (\(f_{\rm{r}}\), \(\kappa\)) and the \(i,j\) spin transition (\(\Omega_{i,j}\), \(\gamma_{i,j}\)). The off-diagonal elements depend on the coupling \(G_{i,j}\) between these two oscillators. In our case, this parameter corresponds to the coupling rate of the spin ensemble to the microwave magnetic field generated by the LER. It is related to the coupling per spin $\bar{g}_{i,j}$ as follows: \(G_{i,j} = \sqrt{\int_{V} \bar{g}_{i,j}^2}\), where \(\bar{g}_{i,j} = \big| \langle \psi_j | \hat{V} | \psi_i \rangle \big| \cdot \Delta P_{i,j}\). Finally, the resonator losses due to its coupling to the experimental set-up (including the transmission line), are parameterized with a complex external loss rate (\(\kappa_{e} e^{i\varphi}\)) that multiplies both the driving and the transmission, and that also accounts for the asymmetries in the transmission signal\autocite{Probst2015}. Using this model, the experimental data are then fitted for the whole frequency and dc magnetic field measurement ranges at once. The fit gives optimal values for $\kappa$, $f_{\rm r}$, $\gamma_{i,j}$, the field-dependent spin transition frequency  \(\Omega_{i,j}(B) \equiv g_{eff}\mu_{\rm B}B + \Omega_{i,j}(0)\), and $G_{i,j}$. The model can be expanded to several spin transitions coupled to the same LER by simply increasing the dimension of the interaction matrix accordingly. This allows characterizing each coupling individually even if the transitions are close to each other in magnetic field (see Fig. \hyperref[fig:3]{3a}).

\setlength{\parindent}{1em}The coupling rate \(G_{ij}\) derived from the fits can be compared to the theoretical values in order to estimate the filling factor of the crystal-LER interaction. The theoretical coupling rate is calculated using the spatial distribution of the microwave magnetic field \(\vec{b}(\vec{r})\) obtained from the LER design. In the simulation, each cell with volume \(\Delta V\) contributes with a coupling \(\tilde{g}_{i,j} (\vec{r}) = \mathcal{A} \rho \Delta V  \big| \langle \psi_j | \hat{V}(\vec{r}) | \psi_i \rangle \big| \cdot \Delta P_{i,j}\), where $\mathcal{A}$ is the isotopic abundance and $\rho$ is the density of spins in the crystal. The theoretical coupling rate is then obtained by simply adding the contribution of every cell in the rf magnetic field simulation: \(G_{i,j} = \sqrt{\sum_{k}{(\tilde{g}_{i,j}^{k})^2}} \).

\setlength{\parindent}{0em} \normalsize {\textbf{Acknowledgements}}

\setlength{\parindent}{0em} This work has been funded by the European Union Horizon 2020 research and innovation programme through FET-OPEN grant FATMOLS-No 862893 and the QUANTERA project SUMO. It was also supported by Spanish Ministry of Science and Innovation under grants PGC2018-094792-B-I00, RT2018-096075-B-C21, PCI2018-093116, PID2019-105552RB-C41 and C-44, PID2020-
115221GB-C41/AEI/10.13039/501100011033, and Grant SEV-2016-0686 (MCIU/AEI/FEDER, UE) and by Novo Nordisk Foundation grant NNF20OC0065610. The SUMO project was also co-funded by the Italian Ministry of University and Research. We also acknowledge financial support from the Gobierno de Aragón grant E09-17R-Q-MAD, from CSIC Research Platform PTI-001, and from ONR-Global through Grant DEFROST N62909-19-1-2053. 

\end{multicols}

\newpage

\begin{center}
\section*{\huge{\textbf{Supplementary Information}}}
\end{center}

\bigskip

\part*{\Large {\textbf{RF magnetic field calculation}}}

\smallskip

\setlength{\parindent}{0em}Performing a comparison between the experimental and the theoretical coupling rates requires knowing with high resolution the rf magnetic field spatial distribution \(\vec{b}(\vec{r})\) created by each resonator. We computed it from $2D$ matrices giving the current distribution in the circuit (see Fig. \hyperref[fig:S1]{S1a}), which can be calculated using the commercial software Sonnet for RF simulations, and then applying the Biot-Savart law: 

\begin{equation}\label{eq:S1}
\vec{b} (\vec{r}) = \frac{\mu_0}{4 \pi} \int_C \frac{\vec{dI} \times \vec{r}}{|\vec{r}|^3}
\end{equation}

\smallskip

This calculation demands a high computational power, taking a long time if performed on a CPU. To boost the calculation speed and hence to increase the resolution of the output rf magnetic field, we have employed our own CUDA code in which the calculation is performed in a NVIDIA graphic card. The discretised $3D$ space above the resonator is divided into several cells. The magnetic field $\vec{b}$ at the $k$-th cell can thus be calculated in parallel (Fig. \hyperref[fig:S1]{S1}). This means that the rf magnetic fields in all cells are simultaneously calculated  using:

\begin{equation}\label{eq:S2}
\vec{b}_{k} (\vec{r}_{k}) = \frac{\mu_0}{4 \pi} \sum_{i,j} \left(\frac{dI_{y}^{i,j}\cdot \Delta r_{z}^{i,j,k}}{|\Delta r^{i,j,k}|^3} - \frac{dI_{x}^{i,j}\cdot \Delta r_{z}^{i,j,k}}{|\Delta r^{i,j,k}|^3} + \frac{dI_{x}^{i,j}\cdot \Delta r_{y}^{i,j,k}}{|\Delta r^{i,j,k}|^3} - \frac{dI_{y}^{i,j}\cdot \Delta r_{z}^{i,j,k}}{|\Delta r^{i,j,k}|^3}\right)
\end{equation}

\smallskip

Where $(dI_{x}, dI_{y})_{i,j}$ is the $2D$ discrete current element obtained from the RF simulation and $\Delta r_{i,j,k}$ is the position of the $k$-cell with respect to the $(i,j)$ current elements (Fig. \hyperref[fig:S1]{S1a}).

\part*{\Large {\textbf{Microwave photon number estimation}}}

\smallskip

\setlength{\parindent}{0em} The number of photons ($n$) in a cavity has been estimated from the measurements shown in Fig. \hyperref[fig:S2]{S2} following the expression:

\begin{equation} \label{eq:S3}
n(P_{\rm{in}}, \omega_{\rm{r}}) = \frac{\kappa_{\rm{e}}}{(\kappa_{i} + \kappa_{\rm{e}})^2} \cdot \frac{P_{\rm{in}}}{\hbar \omega_{\rm{r}}}
\end{equation}

\smallskip

In this expression $P_{\rm{in}}$ is the driving power, $\omega_{\rm{r}}$
is the natural frequency of the resonator in $rad/s$, and  $\kappa_{\rm{e}}$ and $\kappa_{\rm{i}}$ are the external and the internal loss rates of the cavity, respectively.

\part*{\Large {\textbf{Broadband spectroscopy analysis}}}

\smallskip

\setlength{\parindent}{0em} We consider the system sketched in Fig. \hyperref[fig:S3]{S3a}, in 
which a crystal of magnetic molecules interacts with the microwave magnetic field generated by the readout transmission line (Fig. \hyperref[fig:S3]{S3b}). The microwave transmission \(S_{21}\) through this line, shown in Figs 1c. (experiment) and 1d. (simulation), has been normalized as follows:

\begin{equation} \label{eq:S4}
|S_{\rm{21}}(B,f)| = \frac{S_{21}(B_{1} , f) - S_{21}(B_{2} , f)}{S_{21}(B_{2} , f)}
\end{equation}

\smallskip

with \(B_{1} < B_{2}\). For \(B_{1}\) being very close to \(B_{2}\), this normalization approximately corresponds to the derivative of the data with respect to magnetic field. This means that any field-independent signal will be suppressed. In this limit, this derivative is similar to the signal detected by conventional Electron Paramagnetic Resonance (EPR). The data obtained from the calculation are normalised following the same procedure. 
A comparison of experimental and theoretical data at $f$ = 4.05 GHz is shown in Fig. \hyperref[fig:S3]{S3c}.

\newpage

\part*{\Large {\textbf{Time-resolved experiments analysis}}}

\medskip

\setlength{\parindent}{0em} Using the quantum master equation to describe the time dependence of the density matrix $\rho_S$ of a resonator coupled to a transmission line, we arrive at the equation of motion for the expectation value of the operator $\hat{a}$ (Notation: in what follows, all quantities are expectation values unless specified by $\,\hat{}\,\,$):
\begin{equation} \label{eq:S5}
a = \,\underset{S}{\Tr} \hat{a} \rho_S
\,\,\,\,\, \Rightarrow \,\,\,\,\,
\dot{a} = \,\underset{S}{\Tr} \hat{a} \dot{\rho}_S = - \left(i \omega_{\rm r} + \kappa \right) a
\end{equation}
where $\omega_{\rm r} \equiv 2 \pi f_{\rm r}$ is the angular resonance frequency of the cavity and $\kappa$ its decaying rate. Including a driving signal sent via the transmission line, with drive 
frequency $\omega \equiv 2 \pi f$, in the master equation and changing to a reference frame rotating at this frequency ($a \rightarrow \tilde{a}$) gives:
\begin{equation} \label{eq:S6}
\dot{\tilde{a}} = - \left[ i \left( \omega_{\rm r} - \omega \right) + \kappa \right] \tilde{a} - i \sqrt{\kappa_{\rm e} e^{i \varphi}} \tilde{r}_{\rm in}
\end{equation}
where $\tilde{r}_{\rm in}$ is the complex amplitude of the input field. Here, $\kappa \equiv \kappa_{\rm i} + \kappa_{\rm e}$ includes the internal losses of the resonator ($\kappa_{\rm i}$) and its coupling to the transmission line ($\kappa_{\rm e}$). The phase in Eq. (\ref{eq:S6}) models the asymmetry in the line shape of the resonator due to modes in the transmission line. From input-output theory, the transmitted signal is:
\begin{equation} \label{eq:S7}
\tilde{r}_{\rm out} = \tilde{r}_{\rm in} - i \sqrt{\kappa_{\rm e} e^{i \varphi}} \tilde{a}
\end{equation}
Equations (\ref{eq:S6}) and (\ref{eq:S7}) model a dynamical measurement of the resonator with an Arbitrary Waveform Generator ($\tilde{r}_{\rm in}$ can be a function of time). The homodyne signal $S_{21}$ measured with a Vector Network Analyzer is just the steady state transmission (i.e. for $\dot{\tilde{a}} = 0$) with constant $\tilde{r}_{\rm in}$:
\begin{equation} \label{eq:S8}
S_{21} \equiv \left| \dfrac{\tilde{r}_{\rm out}}{\tilde{r}_{\rm in}} \right| = \left| 1 - \dfrac{\kappa_{\rm e} e^{i \varphi}}{i \left( \omega_{\rm r} - \omega \right) + \kappa} \right|
\end{equation}
which is the same as Eq. (\ref{eq:3}) from the main text with $\omega_{\rm r} \rightarrow \Omega$, $\kappa_i \rightarrow \gamma$ and $\kappa_{\rm e} e^{i \varphi} \rightarrow \Gamma$. Notice that, in the main text, all frequencies and loss rates are in Hz, whereas here they are given in rad/s. For coupled resonator - spin ensemble systems, Eqs. (\ref{eq:S6}) and (\ref{eq:S7}) are replaced with:
\begin{equation} \label{eq:S9}
\begin{pmatrix}
\dot{x_1} \\ \dot{x_2}
\end{pmatrix}
=
\begin{pmatrix}
- i \left( \omega_r - \omega \right) - \kappa & - i G \\
- i G & - i \left( \Omega - \omega \right) - \left( \gamma + \Gamma \right)
\end{pmatrix}
\begin{pmatrix}
x_1 \\ x_2
\end{pmatrix}
- i
\begin{pmatrix}
\sqrt{\kappa_{\rm e} e^{i \varphi}} \\ \sqrt{\Gamma}
\end{pmatrix}
\tilde{r}_{\rm in}
\end{equation}
\begin{equation} \label{eq:S10}
\tilde{r}_{\rm out} = \tilde{r}_{\rm in} - i \sqrt{\kappa_{\rm e} e^{i \varphi}} \tilde{x_1} - i \sqrt{\Gamma} x_2
\end{equation}
where $x_1 \equiv \tilde{a}$ for the resonator and $x_2$ for the spin ensemble.

If only the resonator is coupled to the transmission line, then $\Gamma = 0$. And if both systems are in the steady state, we retrieve Eq. (\hyperref[eq:4]{4}) from the main text ($b_{\rm n} \equiv x_{\rm n} / \tilde{r}_{\rm in}$) and $S_{21}$ then reads:
\begin{equation} \label{eq:S11}
S_{21} = \left| 1 - i \sqrt{\kappa_{\rm e} e^{i \varphi}} b_1 \right|
\end{equation}

\newpage

\part*{\Large {\textbf{Tables}}}

\begin{table}[h!]
\label{tab:S1}
\begin{center}
\setlength{\arrayrulewidth}{0.50mm}
\renewcommand{\arraystretch}{1.40}
\begin{tabular}{|c|c|c|c|} 
 \hline
 \rowcolor{black}
 \textcolor{white}{\small \textbf{Isotope}} & \textcolor{white}{\textbf{\(I = 1/2\)}} & \textcolor{white}{\textbf{\(I = 0\)}} & \textcolor{white}{\textbf{\(I = 5/2\)}}\\
 \hline
 \textbf{Transition} & 0 to 3 & 0 to 1 & 0 to 11\\ 
 \textbf{\(G\) (MHz)} & 7.9 & 22.0 & 5.0\\ 
 \textbf{\(\gamma\) (MHz)} & 15.9 & 16.8 & 17.7\\
 \textbf{\small Cooperativity} & 129 & 946 & 46\\
 \hline
\end{tabular}
\caption{{\small \textbf{Table S1. Electronic spin-photon coupling parameters.} This table summarises the spin-photon couplings, the inhomogeneous spin broadening and the cooperativity for the electronic spin transitions involving the ground states of the three different [Yb(trensal)] isotopic derivatives. The resonator has \(f_r\) = 2.93 GHz, the cavity loss rate \(\kappa\) = 16.5 kHz, and the crystal had an $5 \%$ Yb concentration.}}
\end{center}

\end{table}

\begin{table}[h!]
\label{tab:S2}
\begin{center}
\setlength{\arrayrulewidth}{0.50mm}
\renewcommand{\arraystretch}{1.40}
\begin{tabular}{|c|c|c|c|c|c|} 
 \hline
 \rowcolor{black}
 \textcolor{white}{\small \textbf{Concentration}} & \textcolor{white}{\textbf{\(8 \% \)}} & \textcolor{white}{\textbf{\(8 \%\)}} & \textcolor{white}{\textbf{\(8 \%\)}} & \textcolor{white}{\textbf{\(5 \%\)}} & \textcolor{white}{\textbf{\(1.9 \%\)}}\\
 \hline
 \textbf{\(f_r\) (MHz)} & 548.5 & 414.3 & 403.6 & 549.5 & 403.6\\ 
 \textbf{\(G\) (MHz)} & 0.42 & 0.4 & 0.65 & 0.31 & 0.09\\ 
 \textbf{\(\gamma\) (MHz)} & 20.0 & 17.1 & 21.8 & 15.0 & 12.9\\
 \textbf{\small Cooperativity} & 2.1 & 2.5 & 5.7 & 2.7 & 0.2\\
 \hline
\end{tabular}
\caption{{\small \textbf{Table S2. Coupling of the ${5,6}$ electronic 
spin transition in [$^{173}$Yb(trensal)].} This table summarises the spin-photon coupling, the inhomogeneous spin broadening and the cooperativity  of the ${5,6}$ electronic spin transition in 
[$^{173}$Yb(trensal)] measured at low frequencies ($<1$ GHz) for different Yb concentrations. The values are obtained at $T=50$ mK, for which the 
thermal population difference between the two levels involved reaches its maximum value.}}
\end{center}

\end{table}

\begin{table}[h!]
\label{tab:S3}
\setlength{\arrayrulewidth}{0.50mm}
\renewcommand{\arraystretch}{1.40}
\begin{tabular}{|c|c|c|c|c|c|c|c|c|c|c|} 
 \hline
 \rowcolor{black}
 \textcolor{white}{\small \textbf{\(^{173}\)Yb transition}} & \textcolor{white}{\textbf{0 to 1}} & \textcolor{white}{\textbf{0 to 1}} & \textcolor{white}{\textbf{1 to 2}} & \textcolor{white}{\textbf{1 to 2}} & \textcolor{white}{\textbf{1 to 2}} & \textcolor{white}{\textbf{2 to 3}} & \textcolor{white}{\textbf{2 to 3}} & \textcolor{white}{\textbf{2 to 3}} & \textcolor{white}{\textbf{7 to 8}} & \textcolor{white}{\textbf{8 to 9}}\\
 \hline
 \textbf{T (mK)} & 10 & 10 & 10 & 10 & 10 & 50 & 50 & 50 & 70 & 300\\  
 \textbf{\(f_r\) (MHz)} & 549.5 & 548.5 & 414.3 & 403.6 & 403.6 & 414.3 & 403.6 & 403.6 & 414.3 & 414.3\\ 
 \small \textbf{\small Concentration} & 5\(\%\) & 8\(\%\) & 8\(\%\) & 1.9\(\%\) & 8\(\%\) & 8\(\%\) & 1.9\(\%\) & 8\(\%\) & 8\(\%\) & 8\(\%\)\\
 \textbf{\(G\) (MHz)} & 0.10 & 0.13 & 0.24 & 0.07 & 0.43 & 0.11 & 0.03 & 0.21 & 0.12 & 0.05\\ 
 \textbf{\(\gamma\) (MHz)} & 1.9 & 2.1 & 1.6 & 2.0 & 2.3 & 0.6 & 0.8 & 1.1 & 5.3 & 1.2\\
 \textbf{\small Cooperativity} & 2.2 & 2.0 & 9.1 & 0.7 & 24.0 & 4.9 & 0.4 & 11.4 & 0.7 & 0.6\\
 \hline
\end{tabular}
\caption{{\small \textbf{Table S3. Spin-photon couplings for nuclear spin transitions.} This table summarises the spin-photon coupling, 
inhomogeneous broadening and cooperativity for the first three nuclear spin transitions, measured on different cavities and crystals.}}

\end{table}

\newpage

\part*{\Large {\textbf{Figures}}}

\label{fig:S1}
\includegraphics[width=18.0cm]{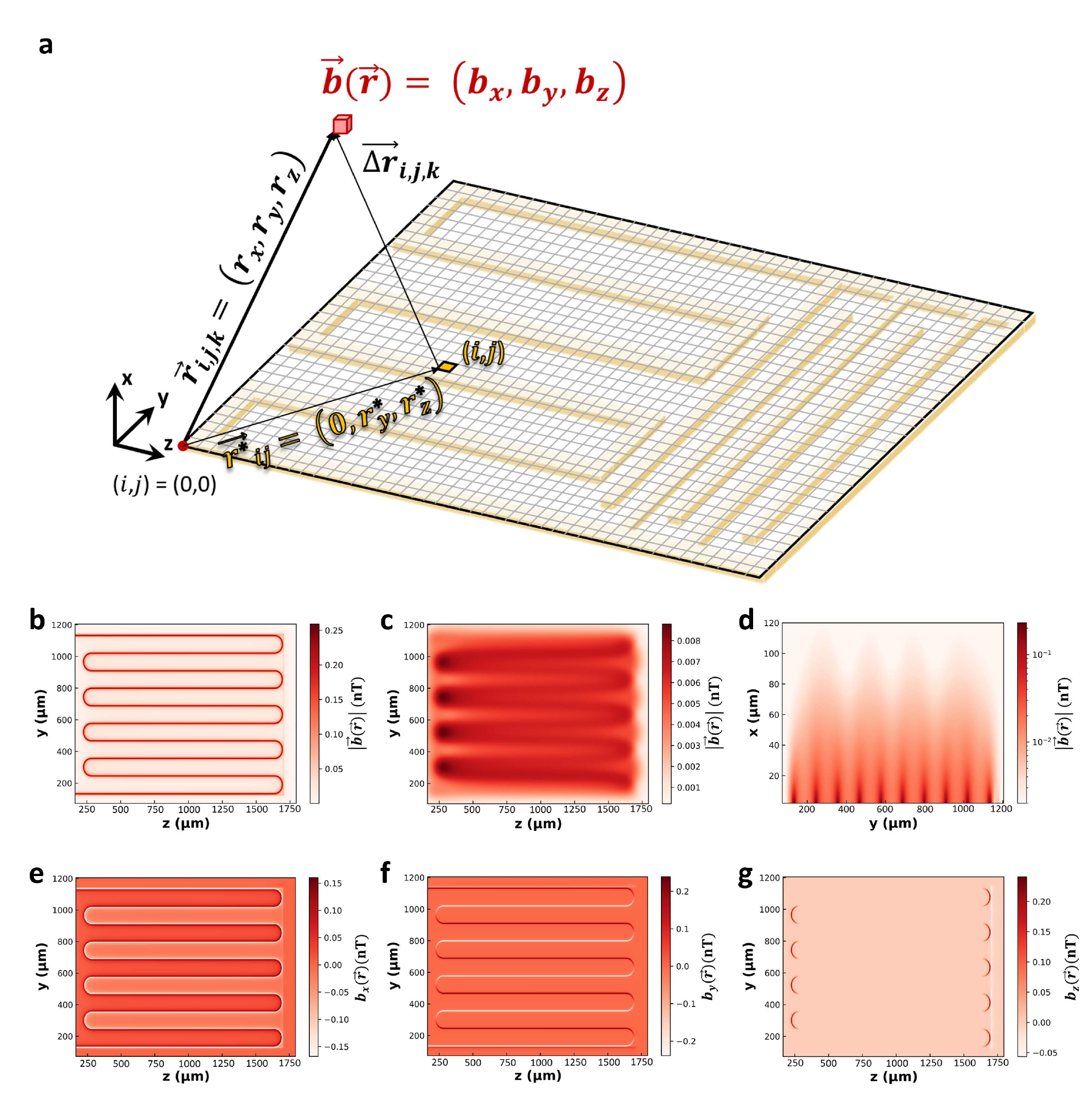}
\captionof{figure}{{\small \textbf{Fig. S1 Calculation of the rf magnetic field (\(\vec{b}(\vec{r})\)) generated  by a superconducting resonator}. \textbf{a} Scheme of the $2D$ discrete current matrix of a resonator. Each matrix element $(i,j)$ contributes to the microwave magnetic field generated at cell $(i,j,k)$. \textbf{b-g} Colour plots of (\(\vec{b}(\vec{r})\)). In these calculations, the resonance frequency of the superconducting cavity is \(f_{\rm r} = \) 549.5 MHz and the driving frequency is in resonance with the cavity. Two $yz$ slices (axes as depicted in (a) and in Fig. 1b of the manuscript) of $|\vec{b}(\vec{r})|$ are presented: \textbf{b} at $x = 1$ $\mu$m, and \textbf{c} at $x = 60$ $\mu$m. \textbf{d} Profile of the rf magnetic field in the $xy$ plane at $z = 1000$ $\mu$m. The $x$, $y$, $z$ components of (\(\vec{b}(\vec{r})\)) are presented in \textbf{e}, \textbf{f} and \textbf{g}, respectively. The microwave magnetic field created by the cavities remains mostly perpendicular to the crystal anisotropy axis, which lies parallel to $z$.}}

\newpage

\label{fig:S2}
\includegraphics[width=18.0cm]{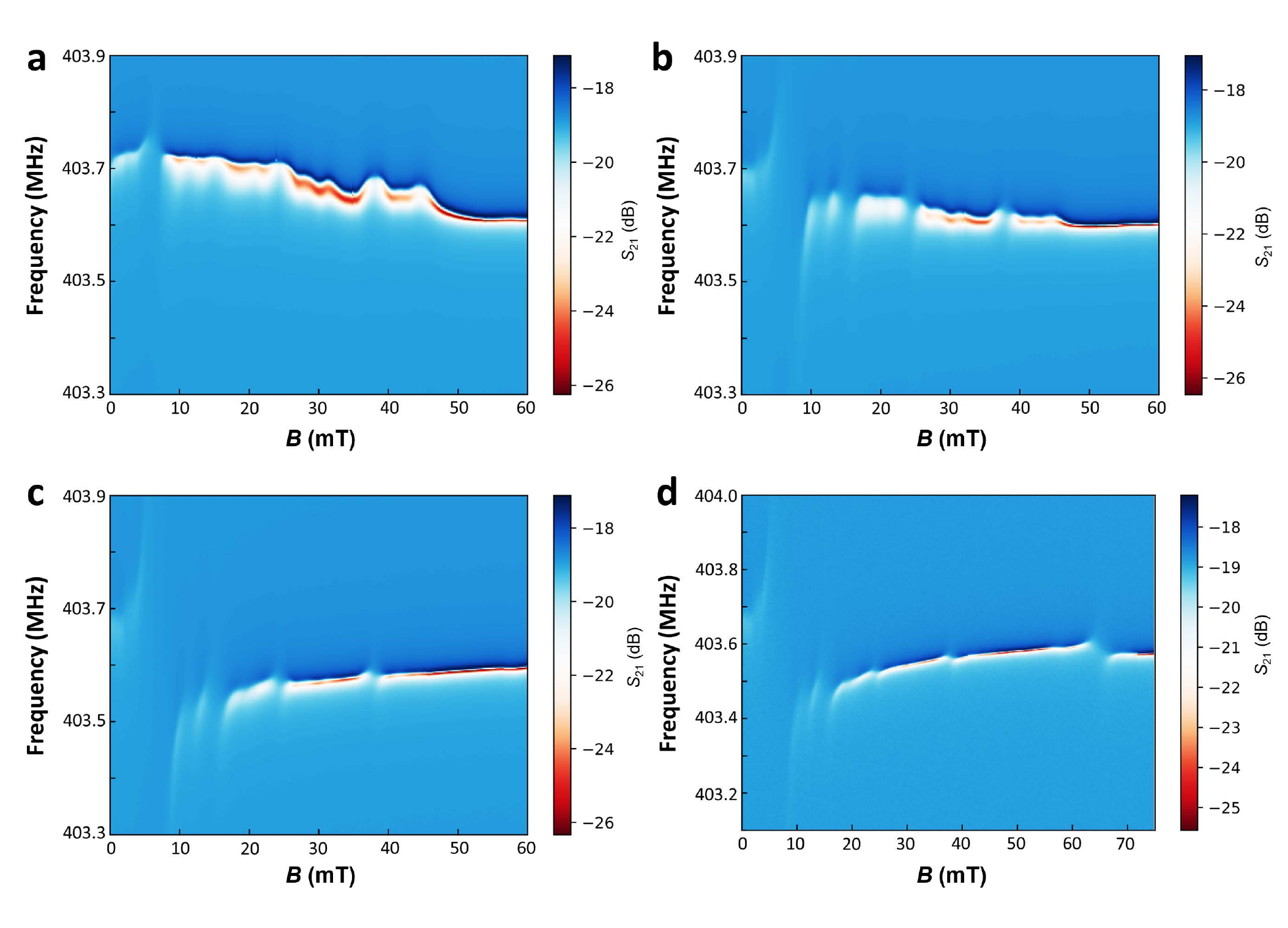}
\captionof{figure}{{\small \textbf{Fig. S2 Optimization of the driving power.} Colour plots of the microwave transmission measured near the resonance frequency $f_{\rm r}$ = 403.6 MHz of a \textit{LC} superconducting resonator as a function of magnetic field. Temperature is $10$ mK for all of them. Measurements at different driving powers are presented, aiming the optimization of photon number in the resonator: \textbf{a}$-50$ dBm ($\sim 10^{13}$ photons), \textbf{b} $-60$ dBm ($\sim 10^{12}$ photons), \textbf{c} $-75$ dBm ($\sim 10^{10}$ photons), and \textbf{d} $-95$ dBm ($\sim 10^8$ photons). Number of photons has been calculated using Eq. (\ref{eq:S3})}}

\newpage

\label{fig:S3}
\includegraphics[width=18.0cm]{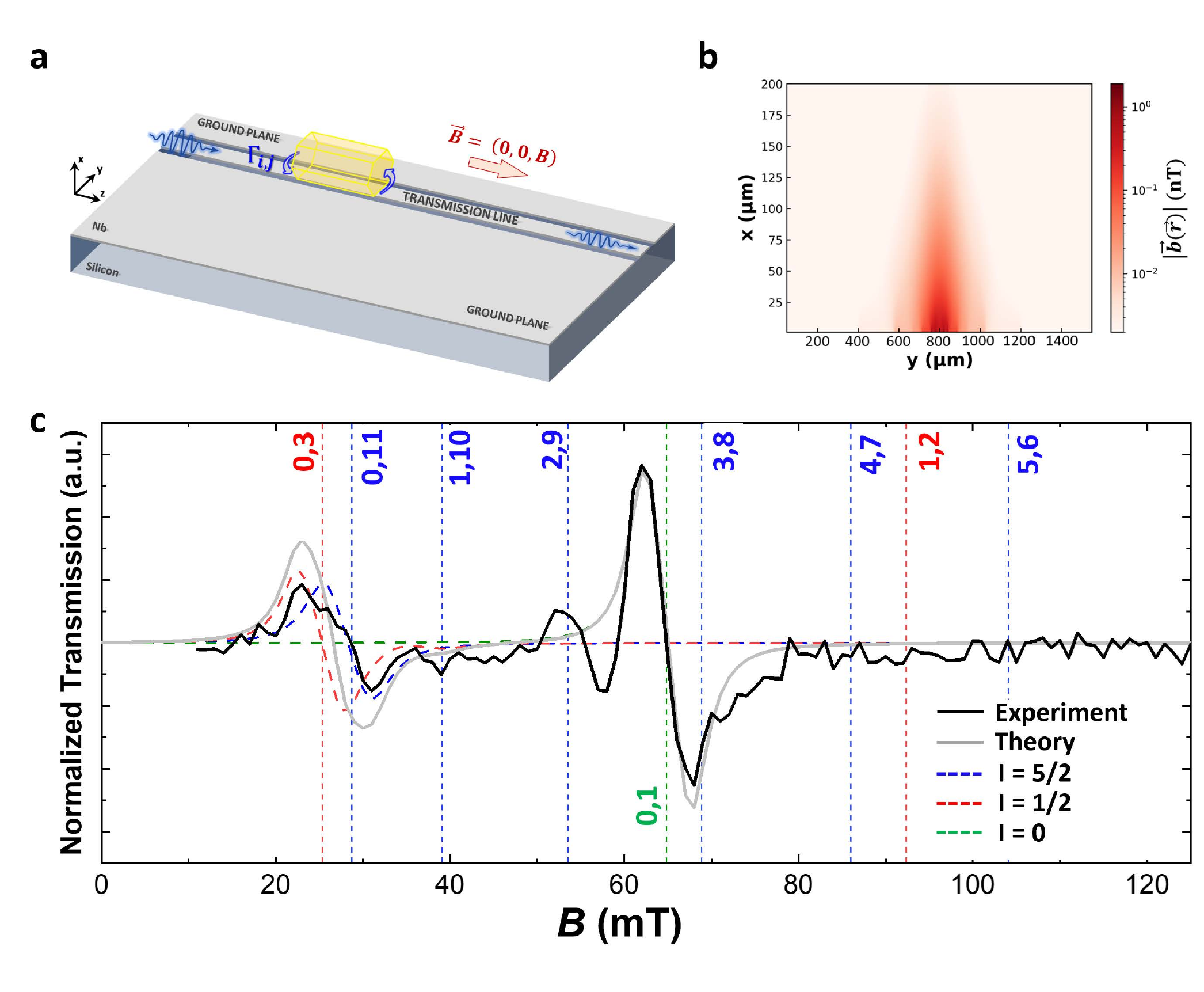}
\captionof{figure}{{\small \textbf{Fig. S3 Broadband spectroscopy experiments. Experimental configuration and measurements.} \textbf{a} Sketch of an [Yb(trensal)] crystal (yellow) placed on a transmission line. Parameter $\Gamma$ is the coupling rate between the transmission line and the spin ensemble (see Eq. (\ref{eq:S9}) in the supplementary information and Eq. (\hyperref[eq:4]{4}) in the main text). The transmission line is parallel to the $Z$ laboratory axis. The external magnetic field lies along this axis, parallel to the crystal \(C_{3}\) anisotropy axis. \textbf{b} Colour plot showing the distribution of the rf magnetic field $\vec{b}(\vec{r})$ generated by the transmission line in the $xy$ plane. \textbf{c} Comparison between experimental (black solid line) and simulated transmission (gray solid line) \(|S_{\rm{21}}(B,f)|\) as a function of external magnetic at $4.05$ GHz and $10$ mK. In this experiment, an [Yb:Lu(trensal)] crystal with a $x$ = 10\(\%\) Yb concentration is directly coupled to the transmission line. All Yb isotopes \(^{171}\)Yb, \(^{173}\)Yb, and the ones having no nuclear spin are present within the crystal in proportions 14\(\%\), 70\(\%\), and 16\(\%\), respectively. The simulated contributions to the transmission signal from each isotope are also included in the image (coloured dotted lines).}}

\newpage

\label{fig:S4}
\includegraphics[width=18.0cm]{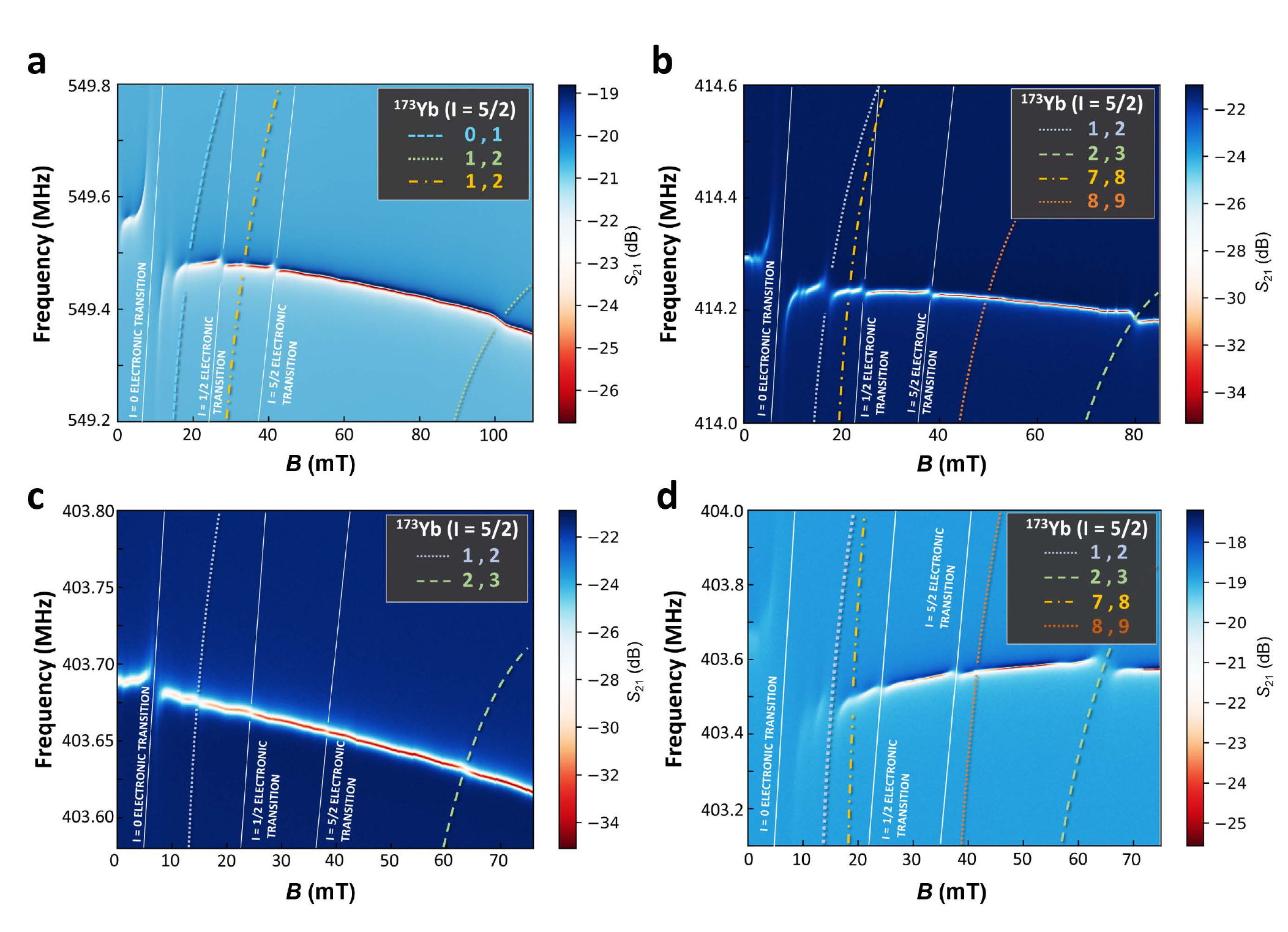}
\captionof{figure}{{\small \textbf{Fig. S4 Coupling $LC$ cavities to nuclear spin transitions.} Microwave transmission \(S_{21}\) as a function of  magnetic field and driving frequency at $10$ mK for: \textbf{a} a \(f_{\rm r}\) = 549.5 MHz superconducting cavity coupled to a crystal with a $x$ = 5\(\%\) concentration. \textbf{b} \(f_{\rm r}\) = 414 MHz cavity coupled to a $x$ = 8\(\%\) crystal. \textbf{c} \(f_{\rm r}\) = 403 MHz cavity coupled to a $x$ = 1.9\(\%\) crystal. \textbf{d} \(f_{\rm r}\) = 403 MHz cavity coupled to a $x$ = 8\(\%\) crystal.}}

\newpage

\label{fig:S5}
\includegraphics[width=18.0cm]{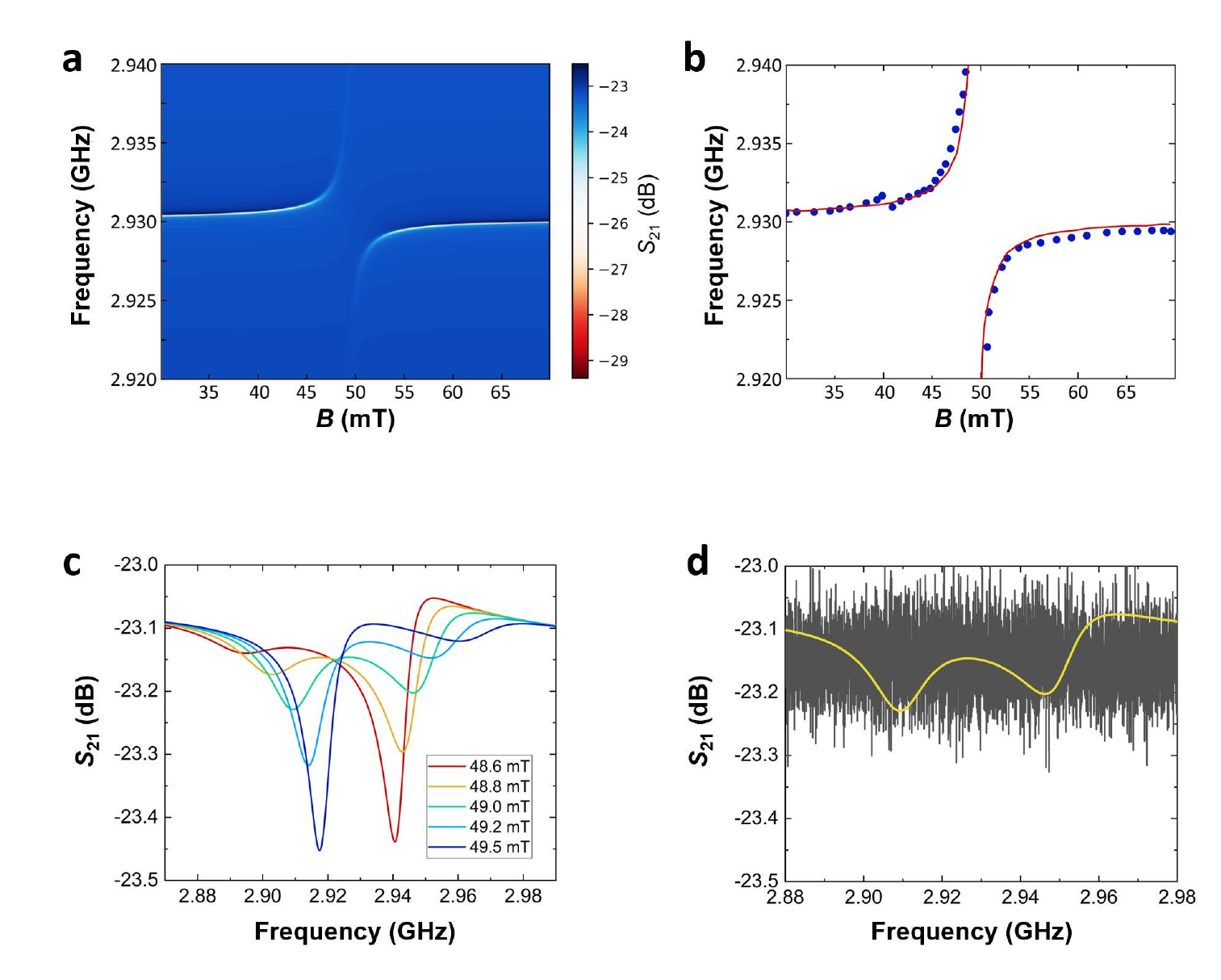}
\captionof{figure}{{\small \textbf{Fig. S5 Strong coupling of cavity photons to the electronic spins of $I=0$ [Yb(trensal)] molecules} \textbf{a} Optimized simulation of microwave transmission \(S_{21}\) as a function of driving frequency and magnetic field. The parameters used in the simulation are optimized through a least squares fit of the experimental data measured at $10$ mK, and shown in Fig. 2a. \textbf{b} Resonance frequency of the cavity-spin ensemble system showing, as a function of magnetic field, the anticrossing that characterizes the strong coupling regime. \textbf{c} Sequence of \(S_{21}\) curves calculated as a function of driving frequency for different magnetic field 
values. In these simulations, taken from (a), a double peak can be observed,
again showing the attainment of the strong coupling regime. \textbf{d} Comparison between experimental and simulated \(S_{21}\) as a function of driving frequency at 49.0 mT. Although a double peak can be observed in the simulation, the experimental noise and the very small visibility make it 
experimentally undetectable.}}

\newpage

\label{fig:S6}
\includegraphics[width=17.5cm]{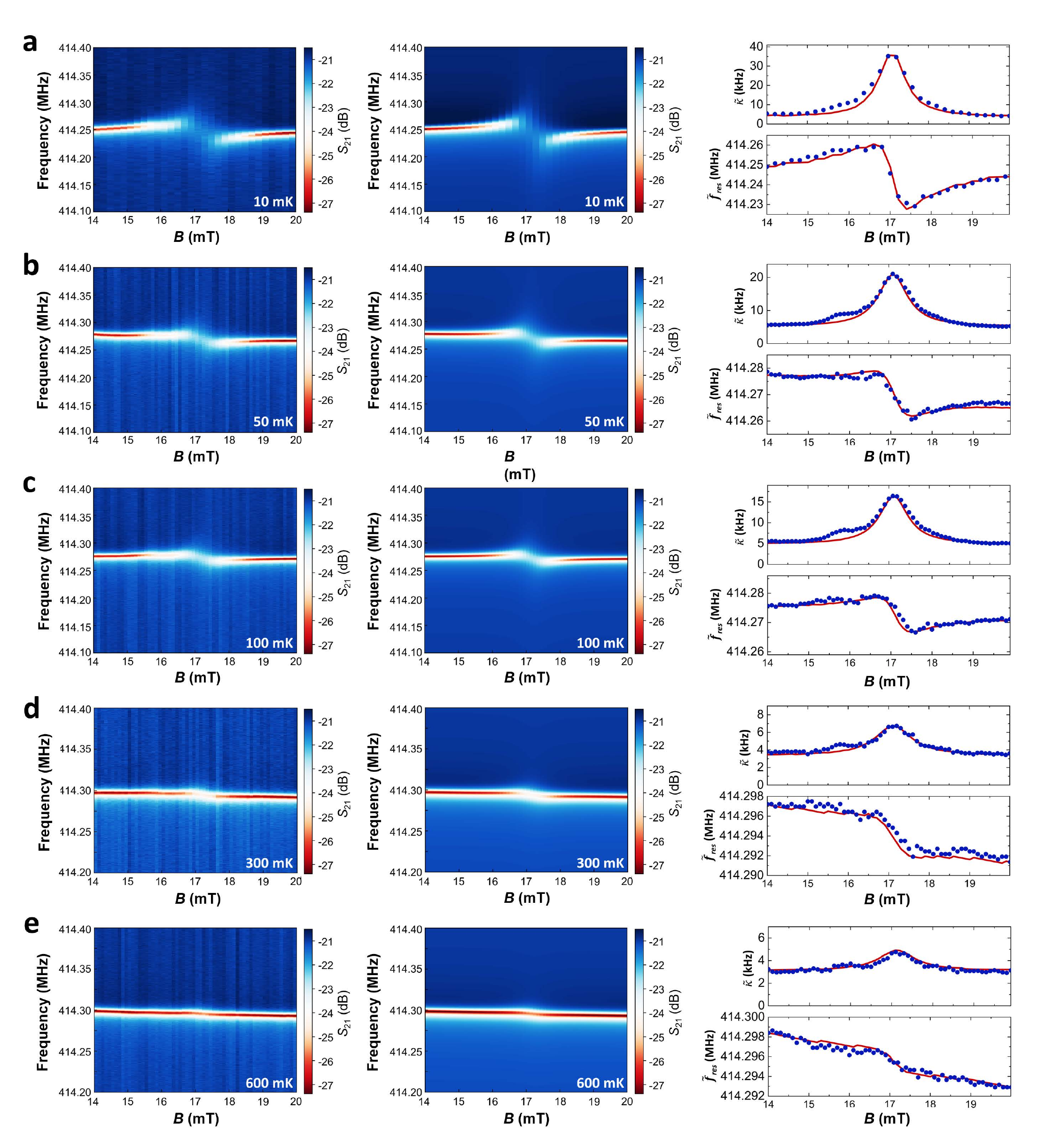}
\captionof{figure}{{\small \textbf{Fig. S6 Evolution with temperature of the transmission measured as a function of magnetic field and driving frequency for a \(f_r\) = 414 MHz superconducting cavity coupled to the ${1,2}$ nuclear spin in transition [\(^{173}\)Yb(trensal)].} Each panel, from {\bf a} to {\bf e}, shows experimental data (left), the simulation obtained with the fitted parameters (center) and the comparison between experimental (blue dots) and simulated data (red line) for \(\bar{\kappa}\) and \(\bar{f}_r\) (right).}}

\newpage

\label{fig:S7}
\includegraphics[width=17.5cm]{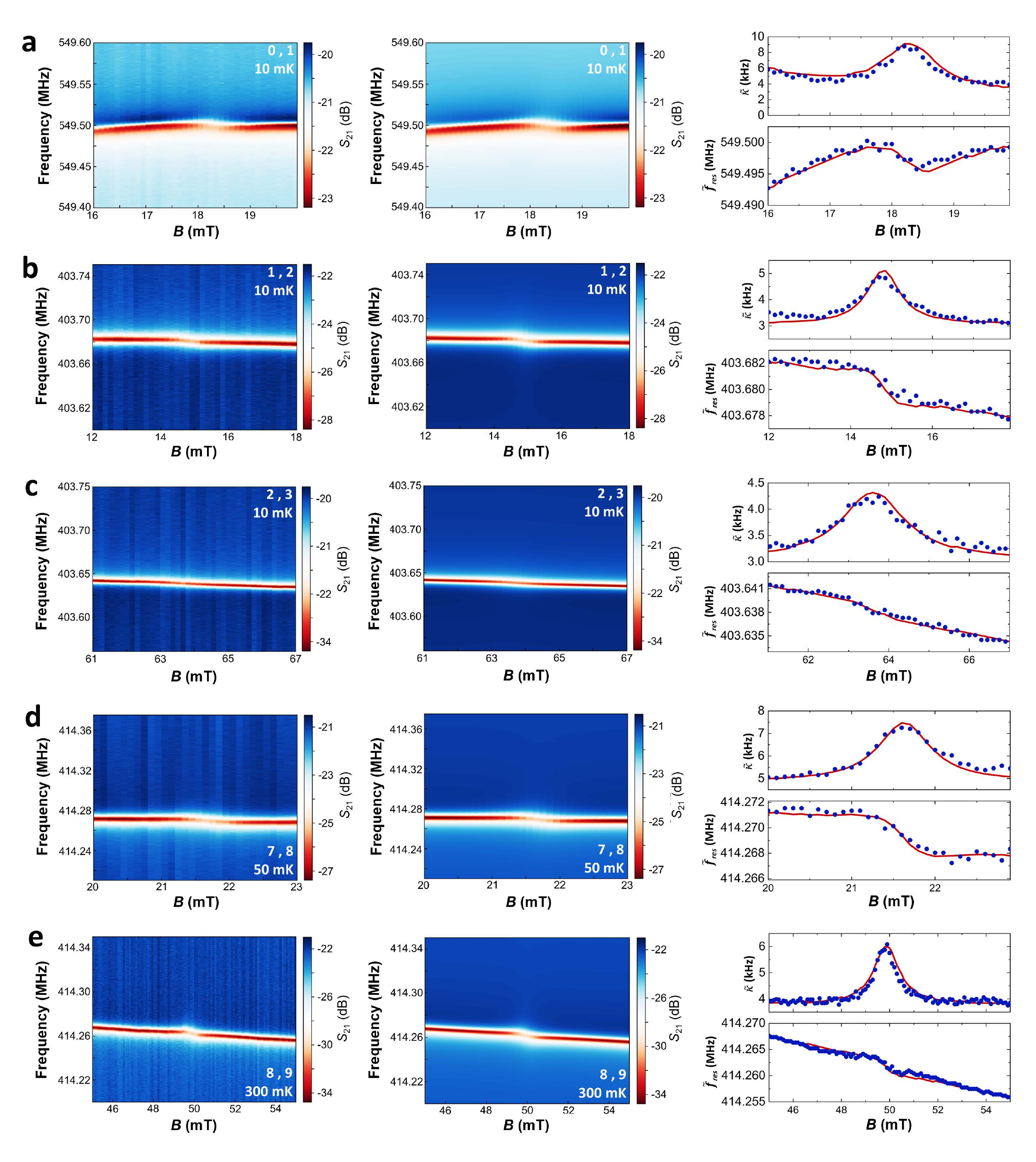}
\captionof{figure}{{\small \textbf{Fig. S7 Additional fits of nuclear spin transitions coupled to different superconducting cavities.}\newline \textbf{a} A \(f_{\rm r}\) = 549.5 MHz cavity coupled to the ${0,1}$ nuclear spin transition in [\(^{173}\)Yb(trensal)] at $10$ mK. The Yb concentration of the crystal was $x$ = 5\(\%\). \textbf{b} A \(f_{\rm r}\) = 403.6 MHz cavity coupled to the ${1,2}$ nuclear spin transition in [\(^{173}\)Yb(trensal)] at $10$ mK. $x$ = 1.9\(\%\). \textbf{c} A \(f_{\rm r}\) = 403.6 MHz cavity coupled to the ${2,3}$ nuclear spin transition in [\(^{173}\)Yb(trensal)] at $10$ mK. $x$ = 1.9\(\%\). \textbf{d} A \(f_{\rm r}\) = 414.3 MHz cavity coupled to the ${7,8}$ nuclear spin transition in [\(^{173}\)Yb(trensal)] at $50$ mK. $x$ = 8\(\%\). \textbf{e} A \(f_{\rm r}\) = 414.3 MHz cavity coupled to the ${8,9}$ nuclear spin transition in [\(^{173}\)Yb(trensal)] at 300 mK. $x$ = 8\(\%\).}}

\newpage

\label{fig:S8}
\includegraphics[width=17.5cm]{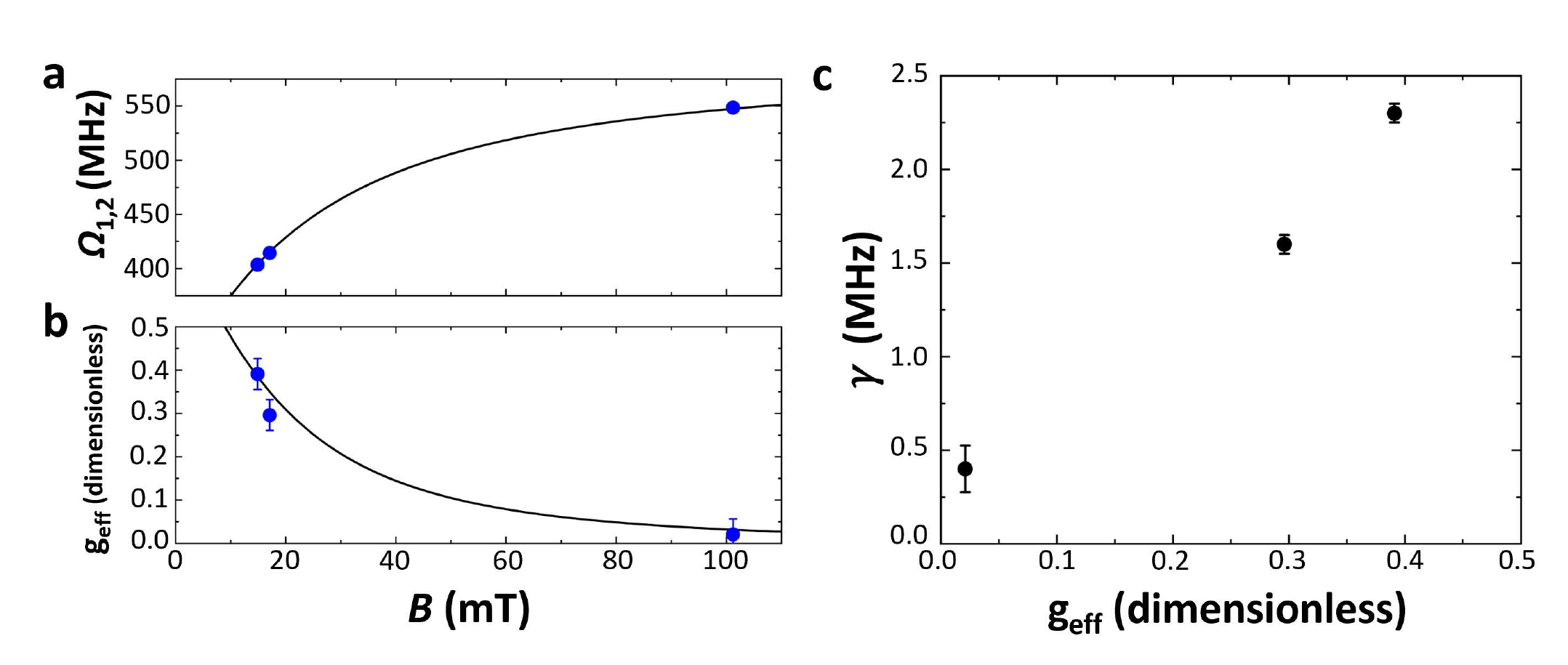}
\captionof{figure}{{\small \textbf{Fig. S8 Variation of the spin line width with magnetic field for the ${1,2}$ nuclear spin transition of [$^{173}$Yb(trensal)].} \textbf{a} Spin transition frequency calculated as a function of magnetic field (full black line) and positions of the specified transition coupled to different resonators (blue dots). All experimental points correspond to crystals with $x$ = 8\%. \textbf{b} Effective slope 
$g_{\rm eff}$ of the nuclear spin transition normalised by $\mu_{\rm{B}}$ ($g_{\rm eff} \equiv \frac{1}{\mu_{\rm{B}}} \cdot \frac{d \Omega_{1,2}}{d B}$) as a function of magnetic field. The line shows the theory and the blue dots are experimental data. \textbf{c} Dependence of the line width on the spin transition slope ($g_{eff}$). The close to linear dependence reveals the direct influence that the inhomogeneous broadening caused by magnetic field fluctuations has on the spin loss rate.}}

\bigskip

\label{fig:S9}
\includegraphics[width=17.5cm]{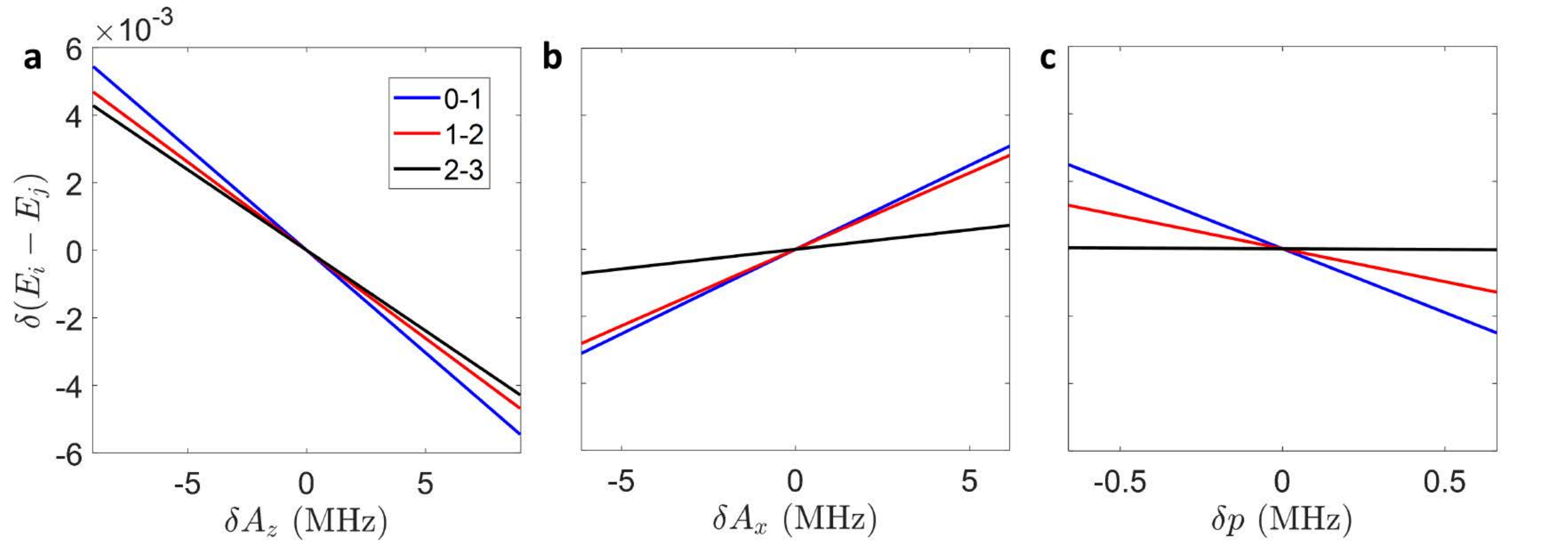}
\captionof{figure}{{\small \textbf{Fig. S9 Effect of the strain in the spin Hamiltonian parameters on nuclear excitations.} Variation of the energy gap of ${0,1}$ (blue), ${1,2}$ (red) and ${2,3}$ (black) nuclear spin transitions (recorded at 20, 17 and 80 mT, respectively) as a function of $\pm 1 \%$ variations in $A_z$ (a), $A_x$ (b) and $p$ (c). The effect of strain in the distribution of the examined parameter on the energy gap is related to the slope of each curve. For all the relevant Hamiltonian parameters (different panels), the ${0,1}$ transition is the most affected, followed by ${1,2}$ and ${2,3}$. This trend is consistent with the observed line widths of 2.1, 1.6 and 0.6 MHz for the three excitations (see main text).}}

\newpage

\label{fig:S10}
\includegraphics[width=17.5cm]{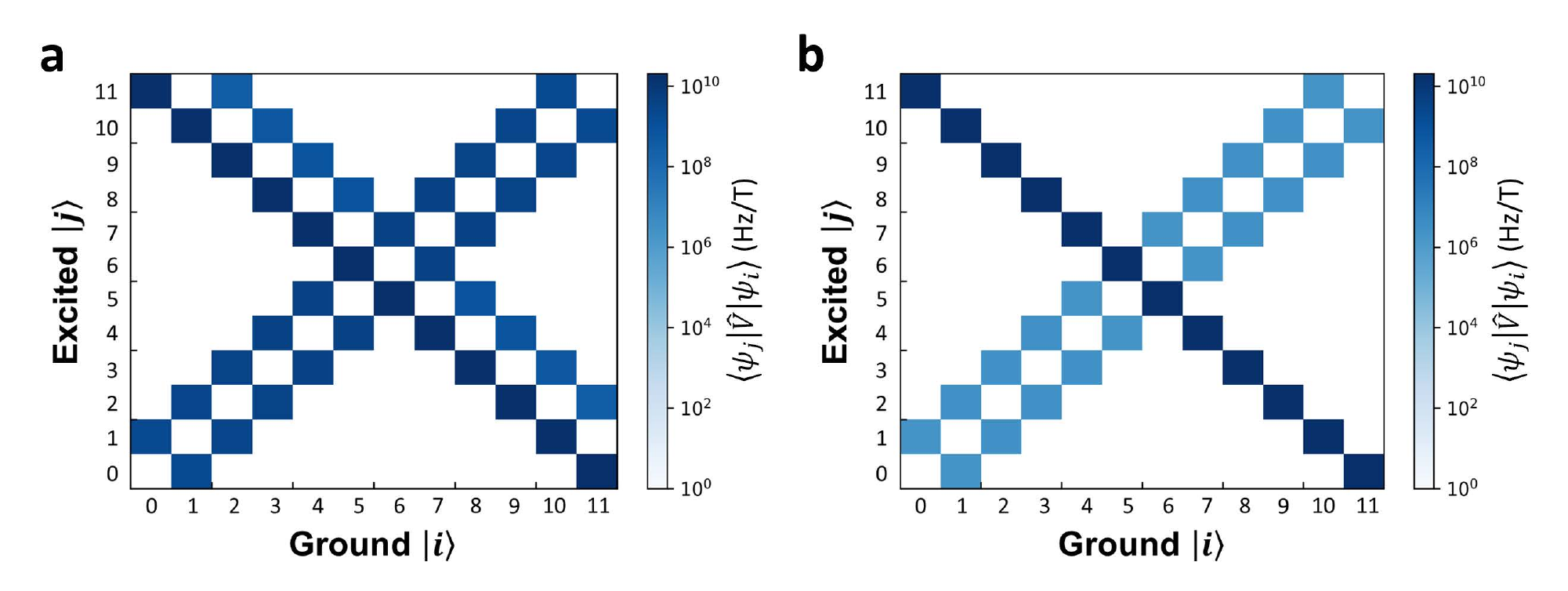}
\captionof{figure}{{\small \textbf{Fig. S10 Matrix elements of spin transitions in [$^{173}$Yb(trensal)].} Calculated matrix elements of all spin transitions at $B$ = 80 mT. Operator $V$ describes the Zeeman coupling of electronic and nuclear spins to the magnetic field created by the cavity: \(V(\vec{r}) = g_{\rm{e}} \mu_{\rm{B}} \vec{b}(\vec{r}) \cdot \vec{S} - g_{\rm{I}} \mu_{\rm{N}} \vec{b}(\vec{r}) \cdot \vec{I}\). In the calculations, \(\vec{b}(\vec{r})\) is normalised and perpendicular to the $C_3$ crystal axis. In panel \textbf{a}, calculations have been performed with the spin Hamiltonian described in the methods section of the main text, which includes a sizeable hyperfine interaction term. In figure \textbf{b}, the hyperfine constants have been set to zero. The matrix elements for the nuclear spin transitions are then reduced by five orders of magnitude.}}

\newpage

\label{fig:S11}
\includegraphics[width=17.0cm]{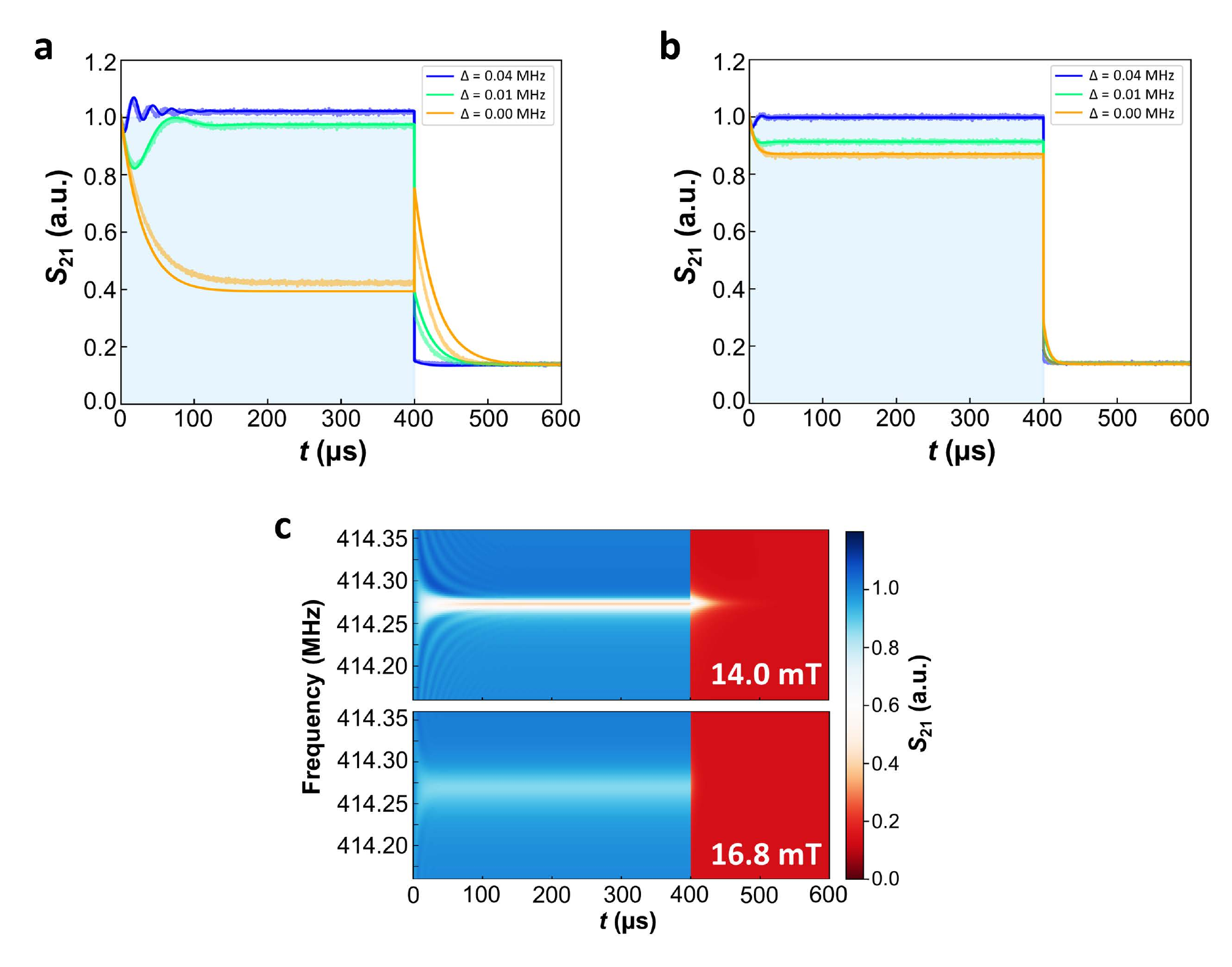}
\captionof{figure}{{\small \textbf{Fig. S11 Dynamic measurements for the ${1,2}$ nuclear spin transition of [\(^{173}\)Yb(trensal)] coupled to a \(f_{\rm{r}}\) = 414 MHz superconducting cavity.} The two panels show data measured at different fixed driving frequencies($f_{\rm{d}}$), and taken from the plots shown in Fig. 4c in the main text: \textbf{a} \(B\) = 14.0 mT, when spins are out-of-resonance with the cavity and \textbf{b} \(B\) = 16.8 mT, when the spins are in resonance with the cavity. The different curves are labelled by the detuning of the driving frequency from the cavity resonance frequency \(\Delta \equiv |f_{\rm{d}} - f_{\rm{r}}| \). Darker continuous lines represent the fits obtained from theory using Eq. (\ref{eq:S9}). \textbf{c} Optimized simulation of the experimental data in Fig.  \hyperref[fig:4]{4c} in the main text. Out-of-resonance (up) and resonant (down) conditions are shown. The transmission has been normalised by the amplitude of the driving pulse (shadowed area in (a) and (b)).}}

\newpage

\label{fig:S12}
\includegraphics[width=17.5cm]{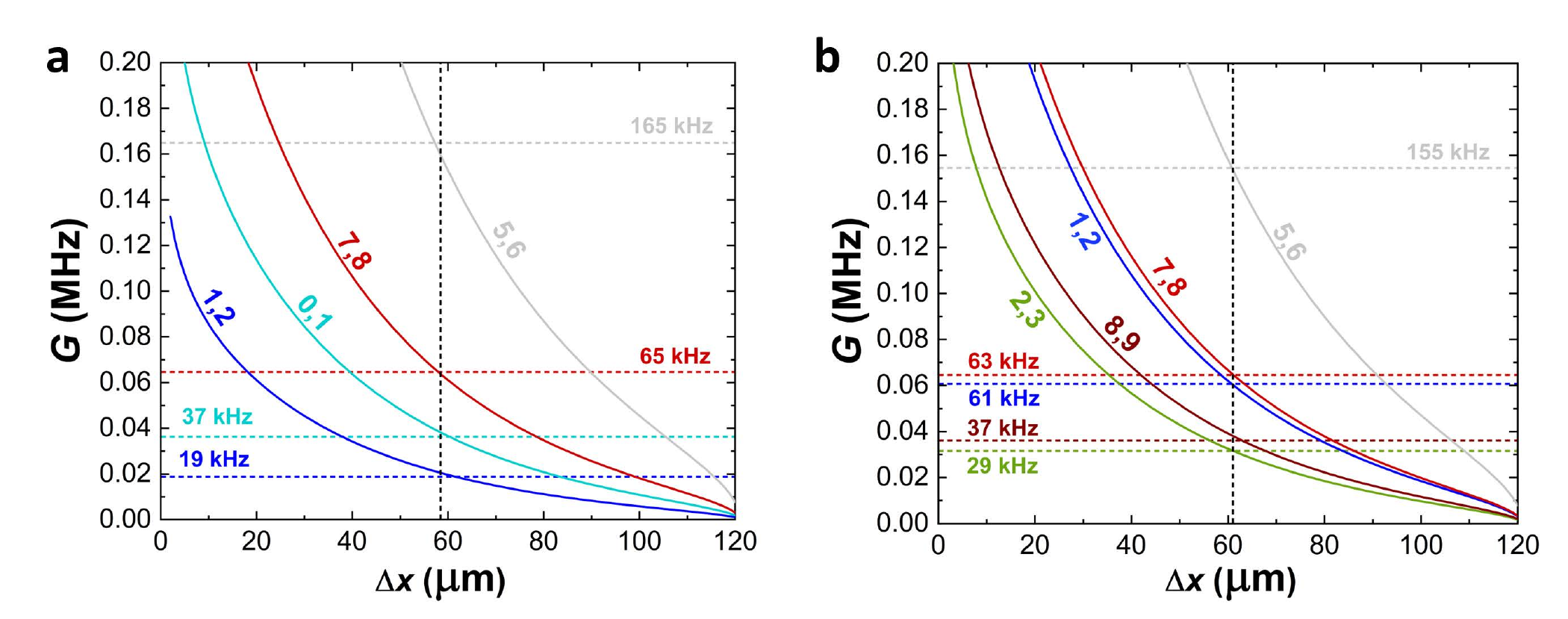}
\captionof{figure}{{\small \textbf{Fig. S12 Estimation of the crystal filling factor.} These figures show the spin-photon coupling calculated for $T=600$ mK as a function of the crystal - cavity distance $\Delta x$ for several [$^{173}$Yb(trensal)] ($i,j$) spin transitions. Dotted lines mark the experimental coupling rates obtained from the experimental data measured at $600$ mK. The plots correspond to two different superconducting cavities. \textbf{a} \(f_{r}\) = 549.5 MHz cavity and a crystal with a $5 \%$ Yb concentration. \textbf{b} \(f_{r}\) = 414.3 MHz cavity and a crystal with a $8 \%$ Yb concentration.}}

\bigskip

\newpage

\bigskip

\section*{\Large \textbf{References}}

\begin{multicols}{2}

\AtNextBibliography{\small}
\printbibliography[heading=none]

\end{multicols}

\end{document}